\documentclass[
aps,
pra,
showpacs
amsmath,
amssymb,
onecolumn,
floatfix,
longbibliography,
letterpaper,
lengthcheck,
superscriptaddress
]{revtex4-2}
\usepackage{svg}
\usepackage{url}
\DeclareUnicodeCharacter{2212}{\textminus}

\usepackage{amsmath}
\usepackage{amssymb}
\usepackage{amsfonts}
\usepackage{bbm}
\usepackage{graphicx}
\usepackage{dcolumn}
\usepackage{bm}
\usepackage[hidelinks]{hyperref}
\definecolor{vastkust}{RGB}{0, 48, 80} 
\definecolor{morkkoppar}{RGB}{0, 108, 92}  
\hypersetup{
  colorlinks=true,
  citecolor=vastkust,
  linkcolor=vastkust,
  urlcolor=vastkust}

\usepackage{float}
\makeatletter
\let\newfloat\newfloat@ltx
\makeatother
\usepackage{algorithm}
\usepackage{algpseudocode}

\makeatletter
\renewcommand{\ALG@name}{Algorithm~}
\makeatother

\usepackage{natbib}
\usepackage[english]{babel}
\usepackage{units}
\usepackage{lipsum}
\usepackage{graphicx}
\usepackage{tikz}
\usepackage[utf8]{inputenc}
\usepackage{pgfplots} 
\usepackage{pgfgantt}
\usepackage{booktabs}
\pgfplotsset{compat=newest} 
\pgfplotsset{plot coordinates/math parser=false}
\usetikzlibrary{arrows}
\usetikzlibrary{angles}
\usetikzlibrary{quotes}
\usetikzlibrary{calc}
\newlength\fwidth
\usepackage{subfigure}
\usepackage{float}
\usepackage{subcaption}
\usepackage{caption}
\usepackage{graphicx}
\graphicspath{{./figures/}}

\usepackage{dirtytalk}
\usepackage{braket}

\usepackage{ragged2e}
\makeatletter
\renewcommand{\@makecaption}[2]{%
  \vskip\abovecaptionskip
  \begingroup
  \justifying
  \small
  \sbox{\@tempboxa}{#1: #2}%
  \ifdim \wd\@tempboxa >\hsize
    #1: #2\par
  \else
    \global \@minipagefalse
    \hb@xt@\hsize{\hfil\box\@tempboxa\hfil}%
  \fi
  \endgroup
  \vskip\belowcaptionskip
}
\makeatother

\begin{document}


\title{Prediction of Molecular Single-Photon Emitters: A Materials-Modelling Approach}

\author{Erik Karlsson Öhman}
\affiliation{Department of Physics, Chalmers University of Technology, 412 96 G\"oteborg, Sweden}
\author{Daqing Wang}
\affiliation{Institute of Applied Physics, University of Bonn, Wegelerstr. 8, 53115 Bonn, Germany}
\author{R. Matthias Geilhufe}
\affiliation{Department of Physics, Chalmers University of Technology, 412 96 G\"oteborg, Sweden}
\author{Christian Sch\"afer}
\email[Electronic address:\;]{christian.schaefer@tuwien.ac.at}
\affiliation{Institute of Applied Physics, TU Wien, Vienna, Austria}

\begin{abstract}
Interfacing light with quantum systems is an integral part of quantum technology, with the most essential building block being single-photon emitters.
Although various platforms exist, each with its individual strengths, molecular emitters boast a unique advantage -- namely the flexibility to tailor their design to fit the requirements of a specific task.
However, the characteristics of the vast space of possible molecular configurations are challenging to understand and explore. Here, we present a theoretical and computational framework to initiate exploration of this vast potential by integrating database analysis with microscopic predictions. Using a model system of dibenzoterrylene in an anthracene host as benchmark, our approach identifies promising new candidates, among them a chiral molecular emitter. Future extensions of our approach integrated with machine learning routines hold the promise of ultimately unlocking the full potential of molecular quantum light-matter interfaces.
\end{abstract}

\date{\today}

\maketitle


\section{Introduction}
Interfacing quantum light with quantum emitters promises new avenues for secure communication\cite{aharonovich2016solid}, efficient simulation of quantum mechanical systems \cite{aspuru2012photonic}, quantum computing \cite{chan2025tailoring}, non-destructive imaging techniques \cite{morris2015imaging}, and novel avenues for material design \cite{baranov2023toward,fojt2024controlling}. 
An ideal single-photon emitter (SPE) delivers single photons of a specific wavelength on demand via a two-level transition with Fourier-limited linewidth.
Moreover, SPEs need to be interfaced to resonant optical structures to enable near-unity-efficiency coupling to a single optical mode \cite{toninelli2021single}, permitting gigahertz operation rates \cite{tomm2021abright}.
Multiple emitters can be cooperatively coupled, promising a flexible handle to modify their emission characteristics.
This might range from enhanced emission speed \cite{nobakht2024cavity}, thus hastening operation times, over topologically reduced emission linewidth \cite{wang2024topological}, the controlled generation of entangled photons \cite{rezai2019polarization}, to the creation of super- and subradiant states of coupled molecules\cite{trebbia2022tailoring,lange2024superradiant}. 

Various platforms have been studied over the last decade.
Among the most promising platforms are semiconductor quantum dots \cite{park2015room}, color centers in solids, such as the widely known nitrogen-vacancy centers in diamond \cite{pezzagna2011creation}, defects in two-dimensional materials \cite{gupta2023single}, and aromatic molecules embedded in organic hosts \cite{toninelli2021single}.
An excellent example of the latter being dibenzoterrylene (DBT) molecules embedded in anthracene \cite{nicolet2007single}.
Single DBT molecules are photostable, feature near lifetime-limited linewidth, have negligible intersystem-crossing yield, and high branching ratio.
Compared to other platforms, molecular SPEs cover a broad range of wavelengths over the visible to near-infrared spectrum \cite{orrit} and feature a unique strength of the flexibility to design a molecule to our liking, i.e., providing an exceptional degree of control over the emitter's characteristics.
However, utilizing this flexibility requires us to understand and identify promising molecules and suitable hosts for such applications \cite{gurlek2025small} -- a highly non-trivial task given the vast space of possible combinations.
\\

\begin{figure*}[ht]
  \centering
  \rotatebox{0}{\includegraphics[width=\textwidth]{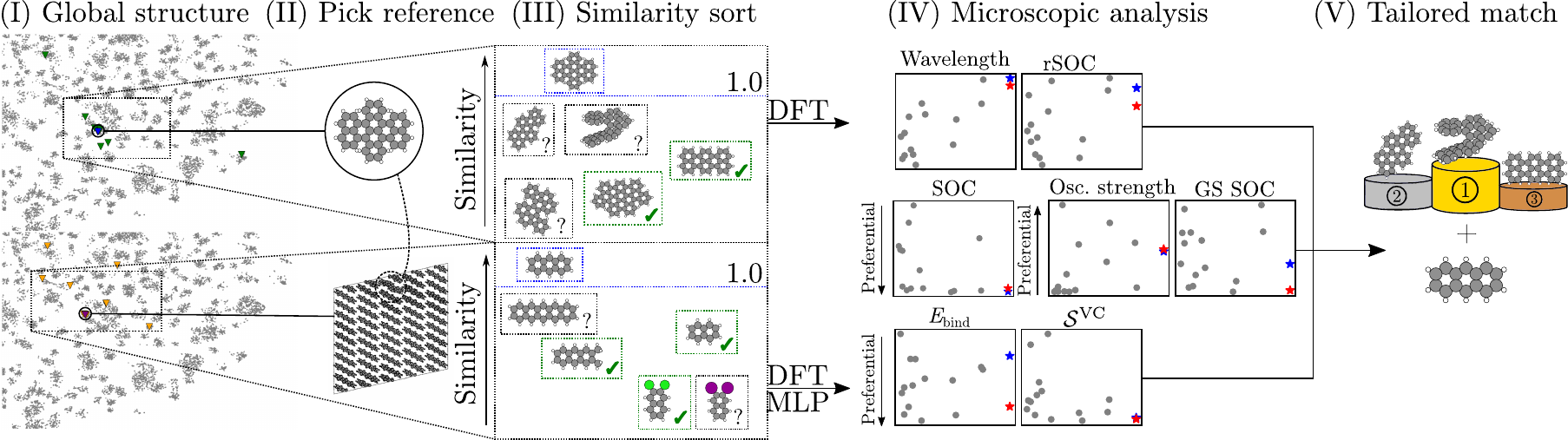}}
  \caption{\textbf{Hunting Emitter-Host pairs:} \textbf{(I)} Entries from the Crystallography Open Database (COD) that are availiable for substructure search by SMILES (approximately $2\cdot10^5$ structures) are collected. The SMILES strings are converted to bitvecors of fixed size and dimensionality reduction and clustering is performed on the dataset by using the t-SNE and HDBSCAN algorithms. \textbf{(II)} A suitable, known emitter-host pair is chosen as a reference. We will limit ourselves here to DBT in anthracene (illustrated), which is among the most studied and promising candidates. \textbf{(III)} The Tanimoto Index is used as a metric and potential replacement emitters and hosts are ranked by their similarity to the references. Suitable high-scoring candidates are examined and selected. \textbf{(IV)} Further microscopic analysis, such as DFT and molecular dynamics calculations, are performed for a set of relevant observables. This includes for example emission wavelength, oscillator strength, various metrics for spin-orbit coupling, vibronic coupling, and formation energies. \textbf{(V)} Results are evaluated and new emitter-host pairs are proposed.
  }
  \label{fig:1}
\end{figure*}

In this work, we describe a first attempt for a fully predictive theoretical approach to scan the vast available configuration space and identify ideal combinations of emitter and host molecules for specific applications. A schematic overview of our approach is illustrated in Fig.~\ref{fig:1}.
We combine database analysis with similarity investigations and microscopic calculations to identify suitable alternatives to established matches, here in particular to DBT in anthracene.
The vast majority of the touched database entries has been experimentally synthesized.
On the methodology side, we incorporate (time-dependent) density-functional theory (TDDFT) and machine learning potentials.
Various promising candidates are identified through our approach.
Among them are new candidates, one being an inherently chiral emitter -- a potential building block for chiral photonics \cite{Forero2025chiral}.
Our approach is validated by the prediction of terrylene, a well known match.
We begin by providing in Sec.~\ref{sec:theory} a brief introduction into required theoretical tools before discussing our results and promising new candidates in Sec.~\ref{sec:results}.

\section{Theory}\label{sec:theory}
Identifying suitable combinations of emitters and hosts can utilize a variety of tools. 
Here, we develop an approach that combines routines from cheminformatics, electronic structure theory, molecular dynamics, and machine learning.
Our brief introduction will not suffice to deliver a comprehensive overview of such a broad set of frameworks.
Readers are therefore encouraged to investigate the cited references for further details.

\subsection{Clustering and Similarity}
In a first step, we apply clustering to explore the global structure of organic molecules with respect to the distribution of known emitters and hosts.
A prerequisite for this is to define a descriptor that uniquely identifies a molecule and a metric which defines a relative distance between two molecules.
We use the SMILES descriptor in combination with Morgan fingerprints and similarity measured by the Tanimoto index, as detailed below.

\subsubsection{SMILES}
\begin{figure*}[!t]  
    \centering
    \subfigure[]{\includegraphics[height=0.275\textwidth]{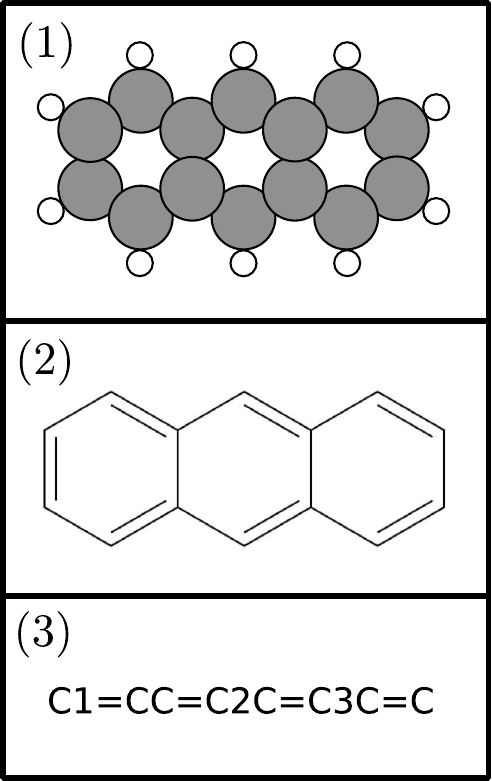}}
    \subfigure[]{\includegraphics[height=0.275\textwidth]{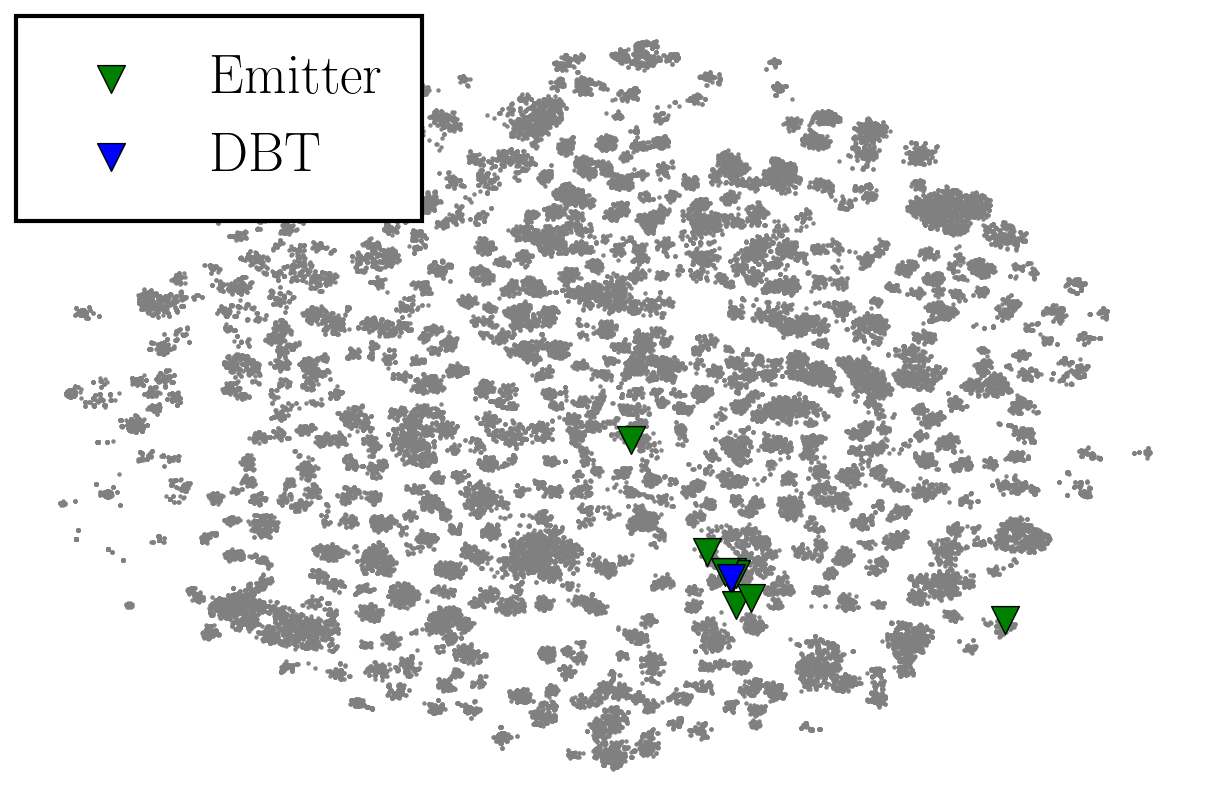}}
    \subfigure[]{\includegraphics[height=0.275\textwidth]{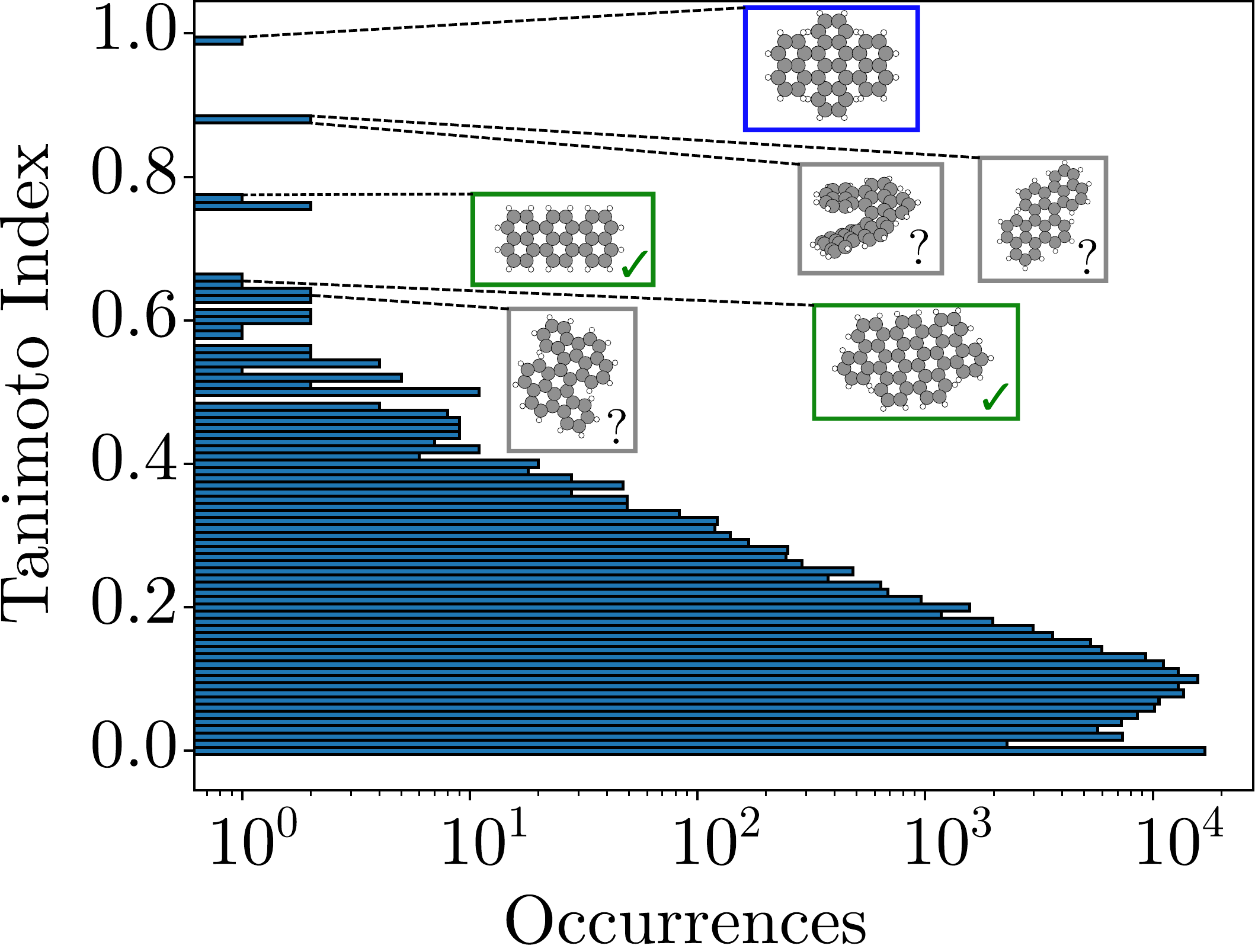}}
    \caption{ \textbf{(a)}  \textbf{Generation of SMILES strings:} Schematic showing the generation of SMILES strings from (1) a molecular structure, which is converted into (2) a graph representing the atomic connectivity, and finally into (3) a linear text-based SMILES representation of the molecule.\textbf{(b)}  \textbf{Global structure of dataset:} t-SNE plot showing the distribution of molecular data. Known emitters are indicated by green triangles, and DBT is highlighted as a blue triangle. \textbf{(c)} \textbf{Similarity Ranking:} Logarithmic distribution of the Tanimoto index of the materials in the Crystallography Open Database with respect to DBT. The insets show relevant samples from the distribution. They are, from top to down and left to right: DBT, terrylene, 4127216, 2000909, 1555531 and BDPB, where the numbers are COD IDs. The blue square denotes the reference material, green squares denote a known emitter, although not necessarily in anthracene, and the grey squares denote previously unknown candidates.}
    \label{fig:smiles_cluster_hist_graphic}
\end{figure*}

Our dataset consists of all entries from the Crystallography Open Database (COD) that are available for substructure search by SMILES \cite{cod_smiles} plus selected known emitter molecules \cite{orrit}  -- totalling 172228 entries at the time of writing. SMILES, or Simplified Molecular Input Line Entry System, characterizes the chemical composition and connectivity of a molecule \cite{smiles}. Atoms and bonds of the molecule are encoded with a series of characters in a string, as illustrated in Fig. \ref{fig:smiles_cluster_hist_graphic} (a). This means that spatial information on, for example, bond distances and angles, is not preserved other than what can be inferred from the abstract chemical bond notation of the molecule.
SMILES are not unique, i.e., one molecule can have two different SMILES, but can be canonicalized \cite{schneider2015get}.
Essentially complementary to SMILES are local descriptors, such as the smooth overlap of atomic positions (SOAP), which provide an explicit representation of atomic coordinates \cite{bartok2013representing}.
Defining simple similarities solely on SOAPs averaged per atom lacks information about the global structure (see App.~\ref{app:soapsimi}).
SOAP (or other local descriptors) encode local information, e.g. about the conjugation system \cite{de2016comparing}, and a fruitful combination with SMILES will be the subject of future research.
\subsubsection{Similarity \& fingerprints}
To be able to quantify the similarity of two molecules, the SMILES strings have to be encoded into a fingerprint, a binary vector of fixed size. We use the open-source Python cheminformatics toolkit RDKit \cite{Rdkit}, and its implementation of the Morgan algorithm \cite{morgan_fp,doi:10.1021/ci100050t}, for this purpose. The creation of the fingerprint depends on two parameters: the bitsize, i.e., the length of the fingerprint that is generated, and the radius, how many neighbors of each atom should be encoded into the fingerprint. For instance, if the radius is set to zero, all atoms are encoded into the fingerprint without any information about their connectivity. Setting the radius to one allows the fingerprint to encode information about each atom and its immediate neighbors, setting the radius to two encodes information about the atom and its two nearest neighbors, and so on. For the purpose of this study we have set the bitsize to 1024 and the radius to two. The similarity of two molecules to each other will be defined via the Tanimoto index $S_{A,B}=c/(a+b-c)$, a well-established metric in the field of cheminformatics \cite{tanimoto}. Here, $a$($b$) is the number of bits set to 1 in fingerprint $A$($B$), and $c$ is the intersection of all bits set to 1 in fingerprint $A$ and $B$. At most $S_{A,B}=1$ for two identical fingerprints and $S_{A,B}=0$ if the fingerprints have no common features. 

\subsubsection{Dimensionality reduction \& clustering}
The general structure of our dataset is revealed using the t-distributed Stochastic Neighbor Embedding (t-SNE) algorithm.
As the name suggests, t-SNE is a stochastic approach to project data onto a low-dimensional representation that is ideal for clustering and visual assessment.
To enhance interpretability, we use the HDBSCAN algorithm to identify clusters and eliminate outliers for ease of visualization.
Both algorithms are implemented through the scikit-learn library and a complexity value of 50 is used for t-SNE and a minimum cluster size of 10 is used for HDBSCAN \cite{scikit-learn}.\\

Systematic structure identification is conceptually similar to identifying a preferred location.
Our global map, (I) in Fig.~\ref{fig:1}, provides a coarse-grained overview of the terrain and allows to separate districts by specific characteristics.
An intuitive approach to finding the ``best SPE in town" is to pick a suitable candidate and begin exploring similar structures, as shown in steps (II) and (III).
Candidates similar enough are then sampled (step IV), i.e., we obtain microscopic information with the tools detailed in the following.
Subsequent ranking brings particularly promising candidates to light (step V). 
First steps in the direction of machine-learning classification techniques (see Sec.~\ref{sec:classification}) pave a way towards more explorative approaches in the future.

%
\subsection{Molecular structure}
Estimating key merits of an SPE, including oscillator strength, transition energy, and intersystem crossing yield, requires a solution of the electronic Schr\"odinger equation.
Density-functional theory (DFT) strikes a balance between accuracy and computational cost and is therefore the predominant approach to describe molecular and crystal structures. The linear many-body Schr\"odinger equation, scaling exponentially in system size, is mapped onto a set of coupled nonlinear single-particle Kohn-Sham equations which can be solved with sub-cubic scaling \cite{kohn1965self}.
Throughout this work, we utilize ORCA \cite{neese2022software} to solve the latter.
For an emitter of interest, we begin with identifying the optimal ground-state configuration by performing geometry relaxation using the 
B3LYP functional \cite{lee1988development,becke1993density} with D4 dispersion correction \cite{caldeweyher2019generally} using the def-TZVPD basis.
Based on those structures, we perform Casida linear-response time-dependent DFT calculations \cite{casida1996time} (including Tamm-Dancoff approximation) to obtain electronic transition strength, frequency, rotary strength, and spin-orbit coupling. Response TDDFT and geometry relaxation are then combined to estimate Stokes-shifts and a figure of merit for vibronic coupling, as detailed in Sec.~\ref{subsec:vib-phon}.

Changes in electronic configuration are coupled with nuclear motion of both the guest molecules and the host crystal. 
Vice versa, thermal occupation of host phonons will result in scattering events and induce homogeneous broadening.
We estimate the molecular motion with the foundation machine-learning potential MACE-OFF \cite{kovacs2023mace} including dispersive interactions. 
This affordable approach allows us (i) to deposit emitter molecules in the chosen host crystal, and (ii) to estimate the coupling strength between local vibrations (to be excited through Franck-Condon transitions) and the delocalized phonons.

\subsubsection{Emitter embedding}
\begin{table*}[!t]
\caption{
\textbf{Microscopic Analysis:} Tanimoto index, oscillator strength, excitation wavelength, absorption circular dichroism, intersystem crossing, reverse intersystem crossing, ground state intersystem crossing, vibronic coupling entropy and binding energy of investigated emitters in anthracene. Details on the two computationally-identified conformers of DBT (\textrm{DBT-CS} and \textrm{DBT-Winged}) are discussed in App.~\ref{app:dbt}. 
The structures of all candidates are illustrated in the Appendix Fig.~\ref{fig:allstructures}.
}
\label{tab:table_emitters_smiles}
\resizebox{\textwidth}{!}{
\begin{tabular}{lccccccccccccc}
\toprule \toprule
 Name/COD ID & Tanimoto Index & $f_{\text{osc}}^{\text{abs}}$ & $f_{\text{osc}}^{\text{em}}$ & $\lambda_{\text{emitter}}^{\text{abs}}$ (nm) & $\lambda_{\text{emitter}}^{\text{em}}$ (nm)&$R\, (10^{40}\text{cgs})$& SOC & rSOC & GS SOC & $\mathcal{S}^{\text{VC}}$ & $E_{\text{bind}}$ (eV) & Refs. \\ \midrule

DBT -- CS\footnotemark[1] & $1.00$ &0.77 &0.77 &674.6& 706.0& -0.17& 0.15 & 4.08 &0.05&0.0061&-0.52&\cite{nicolet2007single}\\

DBT -- Winged\footnotemark[1] & $1.00$ &0.73& 0.73  &707.5&738.6&  0.17& 0.02 & 5.17 &0.68&0.0072&0.07&\cite{DBTwinged}\\
  
Hexa-peri-hexabenzo[7]helicene/4127216 & $0.89$ & 0.13& 0.08 &700.8&1100.7& 276.16 & 0.06 & 5.76 & 1.45&0.0213&-0.09&\cite{4127216}\\
 
Tetrabenzo[\textit{de},\textit{hi},\textit{op},\textit{st}]pentacene/2000909 & $0.88$ & 0.84& 0.88 &594.9&630.3& 73.96  &2.07 &5.02&2.09&0.0115&-0.06&\cite{2000909}\\

Terrylene/-\footnotemark[2] & $0.78$ &1.17&1.24&525.8&551.7& 0.37 &0.05& 2.09 & 0.27&0.0048&-0.62&\cite{10.1063/1.2184311} \\
 
Perylene/-\footnotemark[3] & $0.77$ & 0.50 &0.54& 415.0&447.8& -0.05& 0.0 &0.98 &0.02&0.0110&-0.22&\cite{verhart2016intersystem}\\
 
Benzodiphenanthrobisanthene (BDPB)/-\footnotemark[3] & $0.65$ & 0.79 &0.85& 548.1&589.4&-3.98& 0.15  &5.59&0.59&0.0059&-0.20&\cite{doi:10.1021/jp970894i} \\
 
Diindenozethrene/1555531 & $0.64$ & 0.52&0.03 & 596.0&820.3& 59.20& 0.84 &3.76 & 1.24&0.0726&-0.69&\cite{1555531}\\
 
-/4062916 & $0.61$ & 0.00 &0.00& 326.2&353.1&0.46 & 3.34& 0.56&0.42&0.0279&0.03&\cite{4062916} \\
 
Benzo[\textit{ghi}]perylene/1554212 & $0.60$ & 0.26& 0.32 &371.7&403.0& 4.28 & 0.14 & 0.98&0.15&0.0348&-0.28&\cite{1554212}\\
 
Naphtho[1,2-i]pentahelicene/7155241 & $0.59$ & 0.00 &0.01& 393.4&426.1&5.24& 2.67 &2.20 &1.43&0.0248&0.03&\cite{7155241} \\
 
2.3.7.8-di-(peri-naphthylen)-pyren (DPNP)/-\footnotemark[3] & $0.58$ & 1.46&1.56 &537.6& 562.6&0.75 & 0.07&1.19&0.12&0.0093&-0.53&\cite{LATYCHEVSKAIA2002109} \\
 
Diacenaphtheno[7,8-b:$7^{1}$,$8^{1}$-d]thiophene/2203348 & $0.57$ & 0.19 &0.09& 466.7&597.2& 6.25& 0.22&2.97&1.18&0.0641&-0.30&\cite{2203348}\\
 
-/4118733 & $0.57$ & 0.01 &0.01& 367.9& 399.5&-0.34& 1.59 &1.95&1.99&0.0196&-0.61&\cite{4118733}\\
 
Dibenzo[g,p]chrysene/4107152 & $0.56$ & 0.01&0.02 & 353.1& 386.4& 1.67& 1.63 & 4.73&1.42&0.0206&-0.36&\cite{4107152} \\
 
Pentabenzo[a,d,g,j,m]coronene/4037514 & $0.55$ & 0.00&0.00 & 450.4&482.0&-0.06 & 1.77  &2.59& 0.68&0.0138&0.32&\cite{4037514} \\
 
\bottomrule
\bottomrule
\end{tabular}
}
\footnotetext[1]{Reference emitter.}
\footnotetext[2]{Known emitter in anthracene.}
\footnotetext[3]{Known emitter in other host.}
\end{table*}
The ideal position and orientation of the guest molecule in a host crystal is determined by Gibbs free energy.
Given that most applications are implemented at temperatures near zero Kelvin, we discard the entropic components and deposit the guest molecules such that the formation energy is minimal.
To estimate the formation energy of the guest-host complex, an emitter is randomly inserted into a $5\times5\times5$ anthracene supercell with two anthracene molecules per unit cell. 
We remove then anthracene molecules that would be overlapping with the inserted emitter.
A minimum of 2 and a maximum of 5 molecules were removed when generating the structures. For each emitter and each number of molecules removed, up to 25 structures were generated, or alternatively until 10000 trials had been made.
A detailed algorithm can be found in App.~\ref{app:embedding}.
The combined guest-host complex, as well as the emitter and host individually, are relaxed using MACE-OFF and the formation (or binding) energy is calculated \cite{Huang_2021}
\begin{align}
\begin{split}
    E_{\text{bind}}&=E_{\text{complex}}-E_{\text{emitter}}\\
    &-\frac{N_{\text{atoms}}^{\text{complex}}-N_{\text{atoms}}^{\text{emitter}}}{N_{\text{atoms}}^{\text{supercell}}}E_{\text{supercell}}.
\end{split}
\end{align}
Here, $E_{\text{complex}}$ is the energy of the combined complex (host with embedded molecules), $E_{\text{emitter}}$ is the energy of the emitter only, and $E_{\text{supercell}}$ is the energy of the pure host (full supercell).
The last contribution accounts for the molecules removed from the supercell to provide space for the embedded emitter molecule.
A structure with the lowest formation energy is deemed the most stable configuration of the complex for the specific combination.

\subsubsection{Vibronic coupling}\label{subsec:vib-phon}
A key merit of a SPE is the fraction of photons emitted into the zero-phonon line relative to the overall emission. This quantity, also referred to as the branching ratio, is more intricate in the case of single molecules embedded in a host material. First, the coupling of the electronic transition from $S_1 \to S_0$ with the vibrational modes of the guest molecule gives rise to a series of vibronic zero-phonon-line transitions, described by the canonical Franck-Condon physics. Second, the electronic transition of the guest molecule also couples to the vibrational modes (phonons) of the host material, described by Debye-Waller physics, which leads to phonon sidebands associated with each of the zero-phonon lines. Thus, the branching ratio of the pure electronic component of the $S_1 \to S_0$ transition is diminished by both the coupling to the vibrational modes of the molecule and to the phonons of the host.

The strength of the 0-0-zero-phonon line, i.e., no vibrational and no phononic excitation under emission, can be theoretically estimated from the Huang-Rhys factors of the entire supercell.
This requires (i) calculation of the vibrational modes of the combined system, and (ii) relaxation of the entire supercell using time-dependent DFT calculations for the excited $S_1$ state -- an extremely costly undertaking for large supercells.
In this work, we propose an approximate metric for the branching ratio that is inspired by a combination of molecular Huang-Rhys factor and the probability of scattering in and out of the corresponding mode.
First, the eigenmodes of the emitter-host complex and the eigenmodes of the emitter molecule are calculated with MACE-OFF.

The vibrational mode ($\nu_i$) of the isolated emitter can be projected into the basis of vibrational-phononic modes ($j$) of the entire emitter-host complex, defining the diagonal components of the reduced density-matrix 
\begin{align}
\begin{split}   
\rho_{j}^{\nu_i} &= tr_j(\ket{\nu_{i\text{, emitter}}^{\text{isolated}}}\bra{\nu_{i\text{, emitter}}^{\text{isolated}}})\\
   &= \lvert \braket{\nu_{i\text{, emitter}}^{\text{isolated}}|\nu_{j\text{, emitter}}^{\text{embedded}}}\rvert^2
\end{split}
\end{align}
where the integration appears only over nuclear coordinates associated with the emitter molecule.
The elements $\rho_{j}^{\nu_i}$ range between 0 (no overlap) to 1 for a vibrational mode $v_i$ that is perfectly represented in the complex-mode $j$.
The von-Neumann entropy of this projection 
\begin{align}\label{eq:vNE}
    S^{P}_i&=-\sum_j^{N_{\text{complex}}}\rho_{j}^{\nu_i}\ln{\rho_{j}^{\nu_i}}
\end{align}
provides a measure of how broad the overlap distribution is between local vibrations and the modes of the emitter-host complex.
In other words, $S^{P}_i$ quantifies the degree of mode delocalization~\cite{glaser2024vibrational} and the strength of scattering between the phononic background and local distortions.
Equation~\eqref{eq:vNE} is loosely related to the canonical entropy of the scattering into and out of the vibrational mode $\nu_i$ \footnote{The mode-overlap approximates the matrix element $<i|V_{vib-bare~phonon}|\overline{j}_{bare~phonon}> \approx <i|j>$ in the rate obtained from Fermi's Golden Rule. From the point of view of the vibrational configuration, the electronic transition appears instantaneous and any change in polarization will quickly displace local vibrational and global phononic modes, broadening the typical delta-like transition probability.}.
The vibrational mode of an embedded molecule that is perfectly isolated from the phononic background will feature minimal scattering and $S_i^P = 0$.

Lastly, we need to account for the relative contribution of each molecular eigenmode to the optical emission process, i.e., the vibronic coupling.
We perform excited-state calculations using TDDFT of the emitter and relax their excited-state geometry (resulting in a Stokes shift between absorption and emission wavelength). The relaxed excited state configuration is then inserted into a ground-state DFT calculation to obtain the forces $F$, which are then projected onto the eigenmodes of the isolated emitter
\begin{equation}
    g_i = \lvert\braket{F |\nu^{\text{isolated}}_{\text{emitter, }i}}\rvert^2.
\end{equation}
In other words, a ground-state energy calculation with the relaxed excited-state geometry gives rise to vibrational forces under emission.
Within the double-harmonic approximation, the equivalent vibrational force would be obtained when inserting the $S_0$ geometry into the excited $S_1$ state calculation.
We have decided here to perform the entire relaxation in order to provide emission wavelengths and Huang-Rhys factors for validation.

The weighted entropy, from here on called the vibronic coupling entropy, is then defined as
\begin{equation}
\mathcal{S}^{\text{VC}}=\sum_i^{N^\text{isolated}_\text{modes}} g_i S^P_i.
\end{equation}
A low $\mathcal{S}^{\text{VC}}$ implies therefore that local vibrational modes are only weakly displaced after optical emission, and the modes that are displaced couple weakly to the phononic background.

Appendices~\ref{app:hrentropy-comparison} and~\ref{app:hr-comparison} demonstrate the strong correlation between common Huang-Rhys factors and force projection, the agreement between DFT and MACE-OFF, and the impact of entropic weighting.
Appendix~\ref{app:hrentropy-comparison} provides additional indications why $\mathcal{S}^{\text{VC}}$ is preferable to direct estimates of the Franck-Condon related Huang-Rhys factors.

\subsubsection{Spin transitions}

Intersystem crossing, i.e., the transition between singlet and triplet space, plays a central role in the design of SPEs. Large intersystem crossing combined with a notable triplet lifetime will severely hamper the utilization of the emitter since any transition into the triplet manifold blocks the desired $S_1\leftrightarrow S_0$ emission over the phosphorescence lifetime. On the other hand, an appropiate yield of intersystem crossing is a desired features for quantum memory applications, organic light-emitting diodes \cite{Kreiza2024}, and marker-molecules used on biology  \cite{LI2022214754}.
In this work, we will use the magnitude of spin-orbit coupling (SOC) as a primer for intersystem crossing probabilities. All such SOC calculations are for isolated molecules, but it should be emphasized that the host material can mediate intermolecular intersystem crossing channels, a subject for future studies \cite{kol2005intersystem,10.1063/1.2184311}.

The ideal feature in our approach contains information that correlates with relevant physical process, yet remains computationally simple enough to allow evaluation for many possible guest and host candidates.
We define three different measures for intersystem crossing events, all represented by spin-orbit couplings calculated using TDDFT.
Kasha's rule suggests that intersystem transitions will always appear from the lowest excited state of a spin manifold and our focus will be set on the according spin-orbit couplings.
First, we estimate exothermic transition events from the excited $\ket{S_1}$ singlet state into the manifold of lower energetic triplet states

\begin{equation}
    \text{SOC} = \sqrt{\sum_i} \begin{cases} 
        \lvert \bra{T_i}H_{\text{SO}}\ket{S_1}\rvert^2 & E^{S}_1 >  E^{T}_i  \\
      0 & \text{else}
   \end{cases}.
\end{equation}
This provides a simple figure of merit for energetically favorable, and thus intuitively dominant, transitions. We note that this is a simplified measure and that resonant singlet-to-triplet transitions are more likely than strict transition into the $T_1$ state \cite{Beljonne2001}.

\begin{figure*}[t]  
\centering
\includegraphics[width=\textwidth]{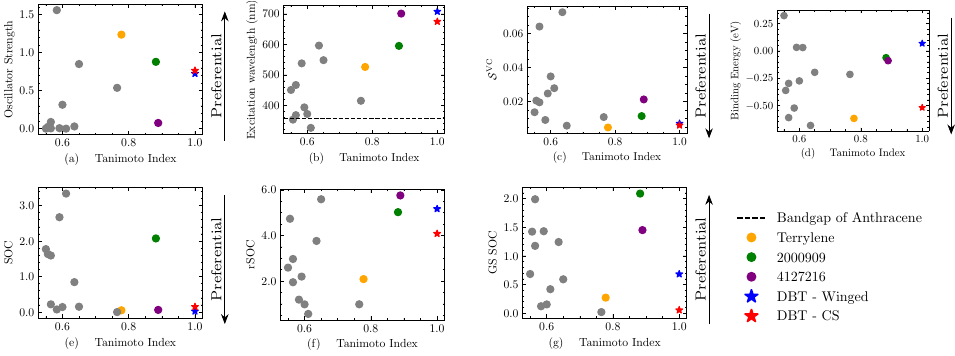}


    
    \caption{\textbf{Microscopic Analysis:} 
    Figures showing \textbf{(a)} the oscillator strength of emission, \textbf{(b)} the excitation wavelength (absorption), \textbf{(c)} the vibronic coupling entropy (with anthracene as host),
    \textbf{(d)} the binding energy (with anthracene as host),  
    \textbf{(e)} the intersystem crossing, \textbf{(f)} the reverse intersystem crossing, and the \textbf{(g)} ground state intersystem crossing plotted for emitters against their Tanimoto index. The reference emitter (DBT) and three especially interesting candidates (terrylene, 2000909, 4127216) are highlighted.}
    \label{fig:analysis}
\end{figure*}
Moreover, the reverse and ground-state SOC may be defined as
\begin{equation}
    \text{rSOC} = \sqrt{\sum_i} \begin{cases} 
        \lvert \bra{T_i}H_{\text{SO}}\ket{S_1}\rvert^2 & E^{S}_1 <  E^{T}_i  \\
      0 & \text{else}
   \end{cases},
\end{equation}

and
\begin{equation}
    \text{gsSOC} = \lvert \bra{T_1}H_{\text{SO}}\ket{S_0}\rvert .
\end{equation}

The rSOC estimates endothermic processes, i.e., transitions into higher-energetic triplet states.
Such processes are deemed dominant in specific guest-host systems \cite{10.1063/1.2184311}.
Lastly, the phosphorescence rate correlates with the SOC from the lowest triplet state into the singlet ground state.

The ideal SPE for photon generation would feature low SOC and rSOC but notable gsSOC.
While none of the three measures is redundant, the SOC is most relevant for SPE applications, and all SOCs tend to be closely linked via the symmetry of the electronic system.

\section{Results and Discussion}\label{sec:results}
The merits defined in the previous section provide us with a set of indicators for the performance of a molecular emitter.
Let us now begin our hunt for new molecular quantum light-matter interfaces.

\subsection{Similiarity}
From Fig. \ref{fig:smiles_cluster_hist_graphic}b we see that most of the known emitters are localized to a single region in the overall parameter space of the dataset. In other words, most known emitters have similar molecular structures. 
This is a first indication that filtering potential emitter candidates by molecular structure could be a meaningful approach. In fact, the notion that similar molecules share similar properties is a cornerstone in the field of cheminformatics wherein it is sometimes referred to as the
\textit{similar property principle} \cite{maggiora}.  
Furthermore, when filtering the dataset with respect to DBT we can see in Fig. \ref{fig:smiles_cluster_hist_graphic}c
 that most structures have a Tanimoto index under 0.4. In fact, $99.9\%$ of structures in our dataset has a Tanimoto index of $0.4$ or lower. Excluding the reference molecule, only $45$ molecules ($0.026\%$) have a Tanimoto index of $0.5$  or higher, and only two of those molecules (COD ID: 2000909, 4127216) have a Tanimoto index over $0.85$. 
 It is therefore sensible to begin the search for new candidates in the vicinity of DBT.
 We would like to add that outliers in Fig. \ref{fig:smiles_cluster_hist_graphic}b suggest that future global optimization strategies should include an explorative component.

\subsection{Characterizing emitter candidates}
The 15 highest-scoring emitters of the similarity analysis are selected for further calculations and evaluations, all structures are visualized in Appendix Fig.~\ref{fig:allstructures}.
A glance over Fig. \ref{fig:analysis} reconfirms that structures with higher similarity ($> 0.8$) have comparable performance while a low Tanimoto Index ($< 0.6$) does not allow any conclusion.
The collected information with molecular identifies and rotary strength can be found in Tab.~\ref{tab:table_emitters_smiles}.
We begin our detailed evaluation with the most fundamental optical features, i.e., brightness and wavelength (Figs. \ref{fig:analysis}a-b).
The excitation wavelengths are positively correlated with the Tanimoto Index since a comparable size of the conjugated $\pi$-system implies a higher similarity in SMILE strings.
Oscillator strength is largely uncorrelated with our chosen similarity measure but tends to be of comparable size within our domain of investigation.
If we compare the highlighted points in Figs. \ref{fig:analysis}a and \ref{fig:analysis}b, we notice a clear correlation: a longer wavelength results in a weaker oscillator strength.

DBT and terrylene defend their place as top emitters via exceptionally small vibronic couplings.
The spin-orbit coupling (SOC), reverse spin-orbit coupling (rSOC) and the ground-state spin-orbit coupling (GS SOC) of the emitter candidates are plotted against their respective Tanimoto indices in Figs. $\ref{fig:analysis}$e-g.
Notice that DBT underperforms in SOC and rSOC, while terrylene is nearly ideal. 
This might allow two different interpretations:
(i) A good emitter should over-perform in a wide range of features, success in only one relevant attribute is insufficient.
(ii) Metrics that utilize the spin-orbit coupling elements are insufficient as reasonable primers for intersystem-crossing rates.
Our goal is therefore to find a suitable trade-off for new candidates, where the relative weight between features depends on the specific task at hand and remains to be identified.
The impact of SOC will be discussed separately for all predictions in order to draw conclusions with and without consideration of SOC. We have observed no clear correlation between emitter size and SOC value.

\subsection{Promising Candidates}
\begin{figure}[t]  
    \centering
    \subfigure[]{\includegraphics[width=0.15\textwidth]{terrylene_go.png}}
    \subfigure[]{\includegraphics[width=0.15\textwidth]{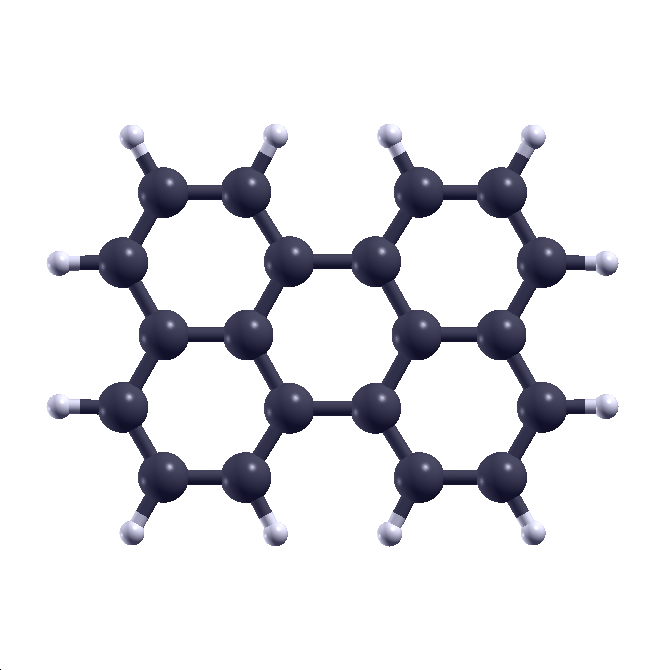}}
    \subfigure[]{\includegraphics[width=0.15\textwidth]{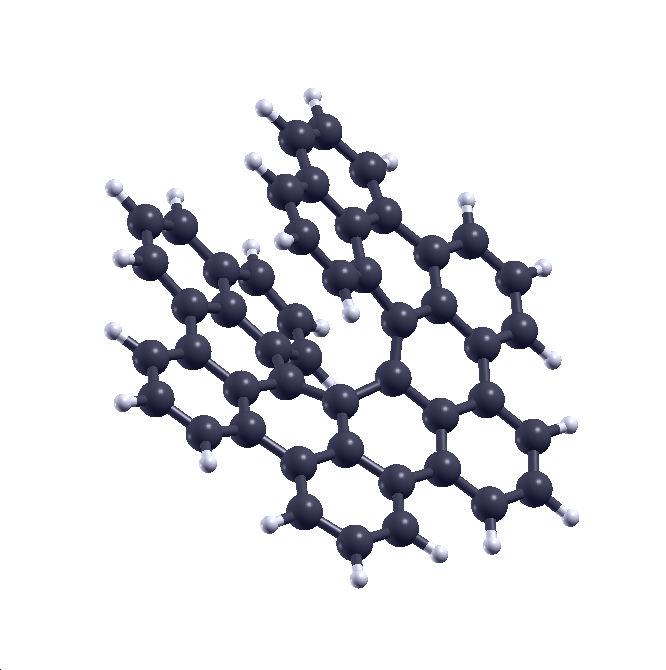}} 
    \subfigure[]{\includegraphics[width=0.15\textwidth]{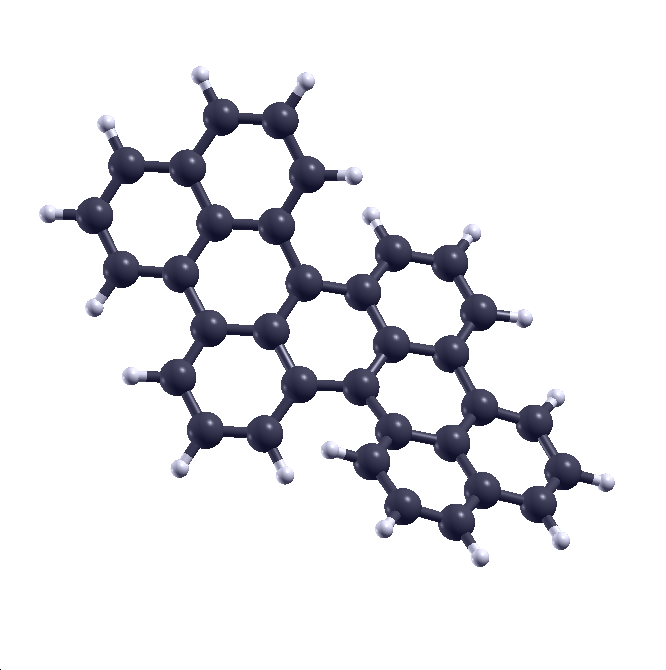}}
    \subfigure[]{\includegraphics[width=0.15\textwidth]{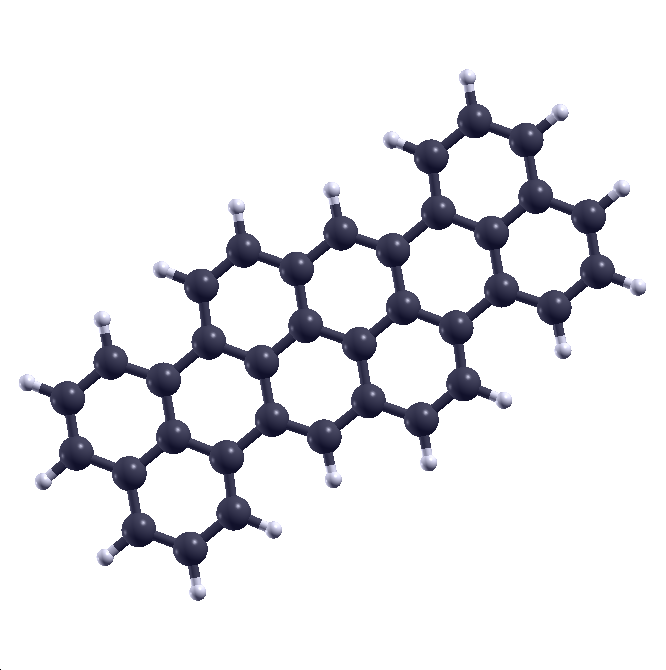}}
    \subfigure[]{\includegraphics[width=0.15\textwidth]{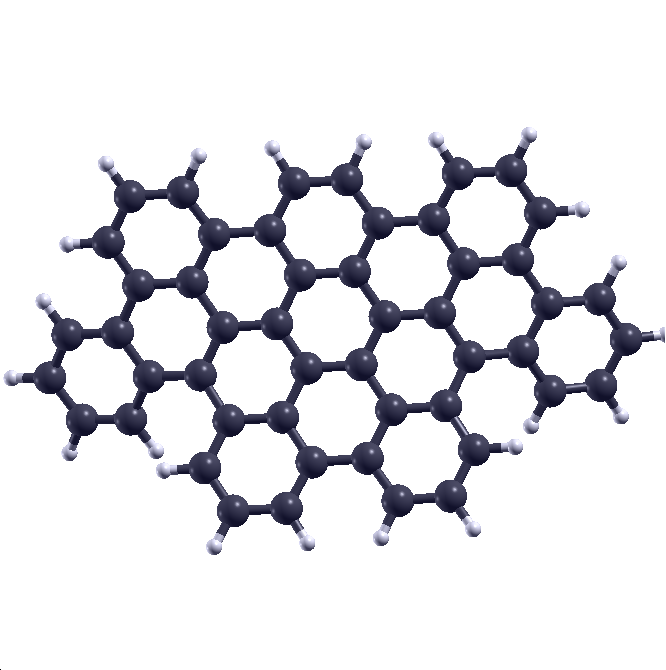}}
    \caption{\textbf{Emitter candidates:}
    \textbf{(a)} The recovery of terrylene serves as validation for our exploitation strategy.
    \textbf{(b)} Perylene illustrates the relevance of intersystem crossing events.
    \textbf{(c)} Hexa-peri-hexabenzo[7]helicene (COD ID: 4127216), is a chiral emitter with decent performance at long wavelength, a promising building block for chiral photonics. 
    \textbf{(d)} Tetrabenzo[de,hi, op,st]pentacene (COD ID: 2000909), follow a similar synthesis route as DBT and presents an ideal validation point for future experimental studies.
    Additional validation candidates:
    \textbf{(e)} DPNP, and
    \textbf{(f)} BDPB. 
    }
    \label{fig:two_plots}
\end{figure}

As a result, we have chosen to highlight six emitters: terrylene and perylene (known in anthracene), 2000909 (unknown), and 4127216 (unknown), BDPB (unknown in anthracene), DPNP (unknown in anthracene), 
all illustrated in Fig.~\ref{fig:two_plots}. 

\subsubsection{Terrylene: A validation}
Experimental research demonstrated that also terrylene performs well in anthracene \cite{10.1063/1.2184311}.
We may thus place extra emphasis on observables for which terrylene and DBT exhibit similar behavior. 
Most notably, as shown in Fig.~\ref{fig:analysis}c, terrylene and DBT have the lowest and second-lowest vibronic coupling entropy among all emitters embedded in anthracene, underlining the relevance of this measure.
Furthermore, terrylene features an exceptionally high oscillator strength.
Overall, our analysis correctly ranks terrylene as one of the best emitters in anthracene.

A second known guest in anthracene is the smaller sized perylene. 
Smaller size reduces its oscillator strength while increasing $\mathcal{S}^{VC}$ -- perylene is therefore clearly a worse SPE than terrylene.
As we shall see in the next section, this moves perylene almost horizontally along the first principal component (PC1) closer to the decision boundary between ``good" and ``bad" emitters.
Perylene and terrylene illustrate therefore nicely the underlying physics, i.e. strong single-photon emission demands foremost a bright and sharp zero-phonon line.
Any coupling to local vibrations or delocalized phonons will negatively impact the performance.
However, more challenging to predict are intersystem crossings mediated by the host crystal.
Direct SOC is vanishingly small in perylene (see Tab.~\ref{tab:table_emitters_smiles}), yet experiments find sizable intersystem crossing which is likely mediated by the anthracene host \cite{WALLA1998117}.
By itself, this results in blinking, i.e., turning the SPE on and off, a feature present for terrylene in anthracene.
In the case of perylene, however, the low GS SOC (see Tab.~\ref{tab:table_emitters_smiles}) traps the system for such an extended time in the triplet manifold that it entirely prevents the utilization of perylene as SPE.



\subsubsection{Exciting candidates}
Out of the small number of investigated emitters, two are deemed especially interesting for further investigation: 2000909 and 4127216, pictured in Fig.~\ref{fig:two_plots}. 
The vibronic coupling entropies of 2000909 and 4127216 are among the lowest of the emitters with transition energy smaller than the bandgap of anthracene.

We can branch our discussion in two possible directions from here on.
First, Tetrabenzo[de,hi,op,st]pentacene (2000909) exhibits an oscillator strength higher than that of DBT at a wavelength domain just between terrylene and DBT -- filling the gap between these two wavelength domains.
The tradeoff are larger SOC elements.
It can be produced analogously to DBT \cite{rauhut1975infrared} and serves therefore as an excellent first test-candidate.

\begin{figure*}[t]
    \centering
    \includegraphics[width=0.473\linewidth]{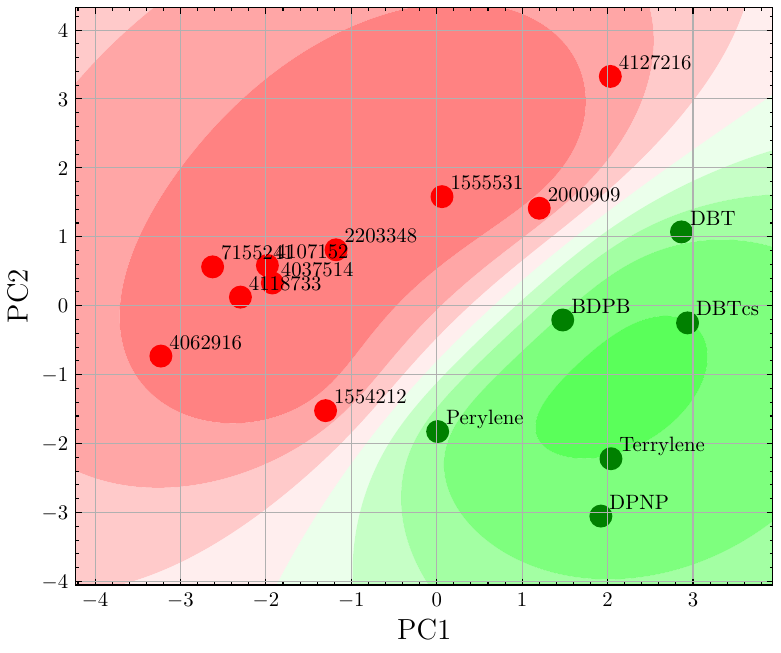}%
    \includegraphics[width=0.527\linewidth]{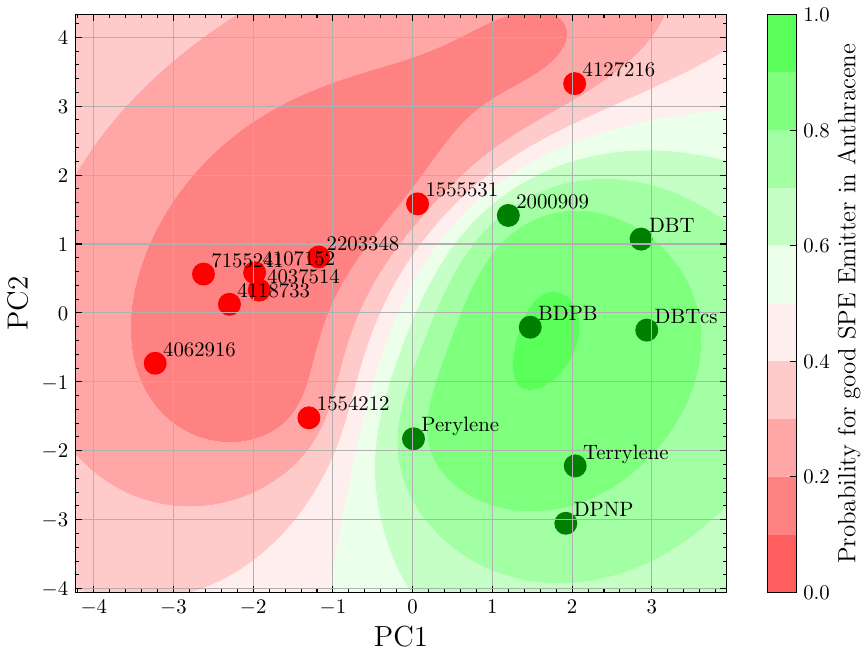}
    \caption{\textbf{Gaussian Process Classification:} 
    The 10-dimensional space \{Tanimoto Index, $f^{\text{abs}}_{\text{osc}}$, $f^{\text{em}}_{\text{osc}}$, $\lambda^{\text{abs}}_{\text{emitter}}$, $\lambda^{\text{em}}_{\text{emitter}}$, SOC, rSOC, GS SOC, $\mathcal{S}^{VC}$, $E_{\text{bind}}$\} (Table~\ref{tab:table_emitters_smiles}) is reduced to the two dominant principal components (PC1 and PC2). An emitter is labeled ``good" if it performs better than average in the most relevant categories (left) $f^{\text{em}}_{\text{osc}} > \overline{f}^{\text{em}}_{\text{osc}}$, $\lambda^{\text{abs}}_{\text{emitter}} > \lambda^{\text{abs}}_{\text{anthracene}}$, $ \mathcal{S}^{\text{VC}} < \overline{\mathcal{S}^{\text{VC}}}$, and $\text{SOC} < \overline{\text{SOC}}$. Right-hand side, the Spin-Orbit Coupling ($\text{SOC}$) is ignored in the labeling, i.e., we do not include any primer for intersystem crossing.
    }
    \label{fig:pca_gpc}
\end{figure*}

The second direction is an entirely new perspective. Hexa-peri-hexabenzo[7]helicen (COD ID:4127216) has a lower oscillator strength with longer wavelength and acceptable SOC couplings.
Controlling the larger vibronic coupling entropy, also indicated by the notable Stark shift, will be more challenging.
However, its emission is strongly chiral (see rotary strength $R$ in Table~\ref{tab:table_emitters_smiles}), which may offer unique opportunities to design inherently chiral SPEs.
Such chiral SPEs would be ideal for near-field chiral sensing and serve as a promising component for chiral photonics \cite{hallett2022engineering,sayrin2015nanophotonic,sollner2015deterministic,zhang2024quantum}.

\subsubsection{Validation candidates}
Other well-tested molecular SPEs are DPNP and BDPB, although no studies have been conducted in anthracene.
Those two candidates might serve as excellent validation cases for our approach, as they can be easily synthesized, are well studied, and their relative performance in experiment would support the theoretical development.

\subsection{Principal Components and Classification}\label{sec:classification}
No emitter is perfect in every metric. 
Which trade-off is ideal depends heavily on the desired task of the emitter.
Central challenges are therefore: 
(i) to identify the relative importance of the various microscopic features (vibronic coupling, spin-orbit coupling, oscillator strength) for a specific task
and (ii) to judge if a candidate might be worth investigating in further detail based on readily accessible information on existing databases.

Our focus in this work is solely on optical coherence.
Four features stand out in their relevance: (i) high oscillator strength, (ii) an absorption wavelength longer than that of anthracene to allow optical excitation of the emitter, (iii) low intersystem crossing, and (iv) a low vibronic coupling entropy.
An emitter that performs better than average in all those criteria will be considered ``good", all others ``bad".
Such \textit{labels} allow the use of machine learning approaches for classification and pattern recognition.
To illustrate the results and understand the inherent relevance of each component, we can calculate the two dominant principal components (PC1 and PC2) from the 10-dimensional space in Tab.~\ref{tab:table_emitters_smiles} (except the rotary strength).
This data, including the above labels color-coded good (green) and bad (red), is visualized in Figs.~\ref{fig:pca_gpc}.
We then apply a Gaussian Process Classifier via the python-package scikit-learn \cite{scikit-learn,williams2006gaussian}.
The kernel is composed of a constant kernel (1.0, 10$^{-3}$ to 10$^{3}$) and a radial basis function (RBF) kernel with length scale (1.0, 10$^{-2}$ to 10$^{2}$).
The Gaussian Process Classifier aims to find an optimal separation into good and bad candidates, drawing a decision boundary between both clusters.
A more intense green (red) color corresponds to a higher probability of a potential candidate being good (bad).
Left (right) side of Fig.~\ref{fig:pca_gpc} includes (excludes) the SOC in the labeling process.
Please note that we train the Gaussian Process Classifier model here with all data points to gather a fundamental understanding. 
Future research will focus on expanding, validating, and testing this model more rigorously.

DBT and terrylene (Fig.~\ref{fig:two_plots}a) are correctly identified as good emitters, consistent with existing experimental observations.
DPNP is identified as another strong candidate, while perylene and BDPB are shifted along $\mathrm{-PC1}$ close to the decision boundary.
Consideration of the SOC degree of freedom has here a substantial impact, with a more restrictive prediction when considering SOC (left).
Host-mediated intersystem crossing events are missing in our current study, resulting in the false labeling of perylene as promising emitter.
The candidates 2000909 and our chiral emitter (4127216) are shifted diagonally along $\mathrm{-PC1+PC2}$ and remain close to the decision boundary.
Especially 2000909 should represent an ideal candidate for experimental validation as its synthesis is similar to DBT.
All other candidates have a low probability to perform on a comparable level.
Our chiral emitter (4127216) is of special interest as it might represent an inherently chiral SPE -- an extremely interesting direction for sensing and chiral photonics.

\section{Conclusion and Outlook}
\label{sec:conclusion_outlook}
Quantum Light-Matter interfaces are a key building block to drive the development of next-generation quantum technology.
Molecules are a promising platform due to their brightness, low inhomogeneous broadening, and low cost. 
Most strikingly, their enormous flexibility allows us to tailor the features of an SPE to a specific task at low cost.
Thus far, this potential has not been leveraged, as a lack of predictive theory and resource intensive experiments prohibited systematic exploration of the available chemical space.\\

Our work tackles this limitation with a combination of database research, microscopic analysis, and initial steps to apply machine learning.
First, we extract SMILES from the Crystallography Open Database and isolate structures similar to DBT, one of the top-performers in anthracene.
Second, various critical features, such as oscillator strength, vibronic coupling, and spin-orbit coupling strength, are calculated using density-functional theory and machine learning potentials.
Lastly, those quantities serve as figures of merit for optical coherence and are then ultimately used to classify possible candidates.\\

The first instance of this methodology identifies up to six alternatives for DBT in anthracene.
First, it correctly predicts terrylene, a known match, which confirms that our predictions are meaningful. 
Among all the emitters considered, Tetrabenzo[de,hi, op,st]pentacene (COD ID: 2000909) performs well above average for all metrics and can be synthesized following a similar route as DBT -- an excellent candidate to validate theory with moderate effort.
Tetrabenzo[de,hi, op,st]pentacene has an excitation wavelength that lies between those of terrylene and DBT, thereby filling the gap in that spectral regime. 
An especially exciting emitter of interest is Hexa-peri-hexabenzo[7]helicene (COD ID: 4127216), which performs comparably to Tetrabenzo[de,hi, op,st]pentacene across all categories, with the exception of oscillator strength. 
Despite its lower oscillator strength, Hexa-peri-hexabenzo[7]helicene merits further investigation due to its distinctly chiral character, which could prove valuable for applications in chiral photonics and chiral near-field sensing~\cite{zhang2024quantum}. 

In addition to the identification of promising molecular SPE candidates, our approach provides additional context to the established empirical guidelines for ``good" SPEs. 
Our figures of merits, e.g., the vibronic coupling entropy, are computationally affordable, and the consistency in preliminary classification underlines their usefulness.
Although additional work will be required for a reliable assessment, our current results support the established development guidelines.

Based on this manuscript, various extensions can be incorporated in the future.
One is to connect our current approach to the true optical coherence of the SPE, i.e., developing a predictive framework for the photon statistics and coherence of the emitter.
This would involve a consistent treatment of the quantum system and structured bath in the presence of nanophotonic elements -- a demanding but reachable goal due to recent progress in the efficient description of many-body system-bath dynamics \cite{muller2024hierarchical,lindel2024quantized}.
Validating the suitability of our predicted chiral SPE could have considerable implications for chiral photonics~\cite{Forero2025chiral,lodahl2017chiral} and chiral sensing.
A second development direction is to establish a global exploration strategy.
We focused on the chemical neighborhood around known matches, an idea loosely based on the \textit{similar property principle} that exploits our existing knowledge.
However, combinations with (multi-fidelity) Bayesian optimization suggest that our technique can be extended with a global, i.e.  explorative, component \cite{snoek2012practical,judge2024applying}.

This manuscript lays the first stone for a predictive design strategy for molecular quantum light-matter interfaces, leveraging their chemical flexibility to identify ideal quantum systems that can be tailored to a wide range of tasks.

\begin{acknowledgments}
The authors acknowledge funding from the Nano Area of Advance, an Excellence Cluster of Chalmers.
C.S. acknowledges funding from the Horizon Europe research and innovation program of the European Union under the Marie Sk{\l}odowska-Curie grant agreement no.\ 101065117. R.M.G. acknowledges funding from the Swedish Research Council (VR starting Grant No. 2022-03350), the Olle Engkvist Foundation (Grant No. 229-0443), the Royal Physiographic Society in Lund (Horisont), and the Knut and Alice Wallenberg Foundation (Grant No. 2023.0087). D.W. acknowledges financial support from Germany’s Excellence Strategy - Cluster of Excellence Matter and Light for Quantum Computing (ML4Q) EXC 2004/1-390534769 and the European Union (ERC, MSpin, 101077866).
\end{acknowledgments}

\appendix

\begin{figure*}[t]
    \centering
       \subfigure[\label{fig:sumg_vs_entropy}]{\includegraphics[height=0.25\textwidth]{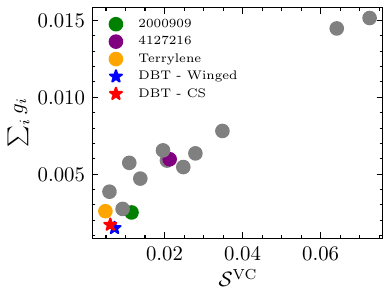}}
        \subfigure[ \label{fig:SumHRvsS}]{\includegraphics[height=0.25\textwidth]{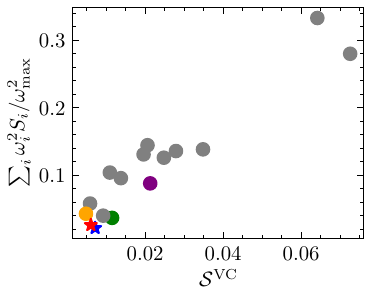}}
     \subfigure[\label{fig:sumdirectFCvibtosuperlattice}
]{\includegraphics[height=0.25\textwidth]{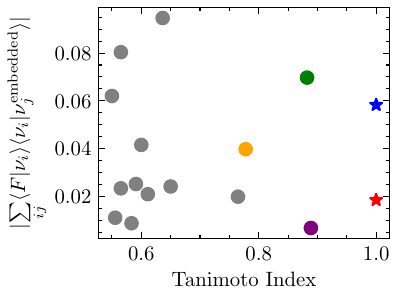}}
    \caption{\textbf{(a) Sum over emission-force projected on vibrational modes vs vibronic coupling entropy:}
    A clear positive correlation is found with only small variations due to the coupling between local vibrations and global phonons. \textbf{(b) Weighted Huang-Rhys sum vs vibronic coupling entropy:}
    A clear positive correlation is found.
    This comes as no surprise as the entropy shows a clear correlation with the sum of the projected forces (see Fig.~\ref{fig:sumg_vs_entropy}).
    The entropy includes, however, an additional estimate of the coupling between local vibrational and global phononic modes, thus accounting for scattering events that dictate temperature progression (see also Fig.~\ref{fig:SumHRvsSvsSumg}).  \textbf{(c) Absolute of the sum over FC vibronic coupling with combined system:}
    The Franck-Condon vibronic coupling, i.e., the vibrational force due to electronic transition, is projected directly onto the entire supercell.
    This is closely related to the Huang-Rhys factor of the combined guest-host system if we ignore the displacement of the phonons due to the charge-displacement of the molecule. In other words, the forces include only molecular contributions that scatter with the phonons of the superlattice.
    The smaller the value, the stronger the 0-0 phonon-line and the more useful the molecular guest as SPE.
    See text for further details.
    }
    \label{fig:HR}
\end{figure*}

\section{Comparison of Huang-Rhys factor and vibronic coupling entropy $\mathcal{S}^{\text{VC}}$}\label{app:hrentropy-comparison}
\begin{figure*}[!t]  
\centering
\subfigure[]{\includegraphics[width=0.31\textwidth]{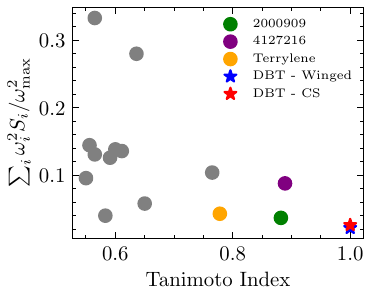}}
    \subfigure[]{\includegraphics[width=0.32\textwidth]{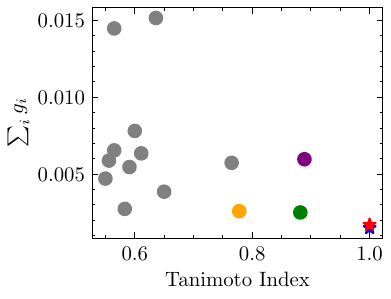}}
    \subfigure[]{\includegraphics[width=0.31\textwidth]{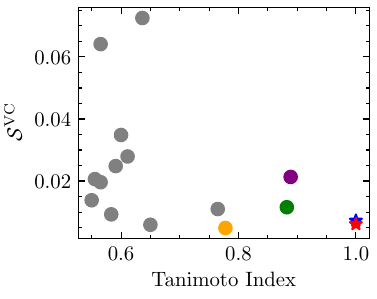}}
    \caption{
    \textbf{Comparison for vibronic coupling estimates}:
    \textbf{(a)} Sum over weighted Huang-Rhys factors, \textbf{(b)} sum over emission-force projected onto vibrational modes, 
    \textbf{(c)} vibronic coupling entropy.
    All three quantities are well correlated, with a slightly larger change between (b) and (c), where local-to-phonon-coupling is incorporated.
    Local vibronic features seem thus more relevant for the overall performance compared to phononic effects, while the latter will impact especially the temperature sensitivity and homogeneous linewidth broadening.}
    \label{fig:SumHRvsSvsSumg}
\end{figure*}

The local vibrational contribution to the vibrational coupling entropy is closely related to the Huang-Rhys factor, supporting the evaluation in this manuscript.
In Fig.~\ref{fig:sumg_vs_entropy}, a clear positive correlation is observed between the sum over the emission-forces projected onto the vibrational modes and the vibrational entropy. 
Fig.~\ref{fig:SumHRvsS} shows a clear correlation between the sum of the Huang-Rhys factors and the vibronic coupling entropy with a high resemblance to Fig.~\ref{fig:sumg_vs_entropy}.
Coupling between local vibrational and global phononic modes plays a subordinate role in the entropy but provides valuable information on temperature-sensitive homogeneous broadening due to phonon scattering.
Those effects result in slight shifts among the datapoints and ensure DBT and terrylene are the best performing guests in anthracene (see Fig.~\ref{fig:SumHRvsSvsSumg}).

We present in Fig.~\ref{fig:sumdirectFCvibtosuperlattice} an alternative figure of merit.
Focusing solely on the Franck-Condon contribution of the molecule and its coupling to the extended supercell, we can estimate the strength of the 0-0 phonon line as
\begin{align}
    \left \lvert \sum \limits_{j} \braket{F|\nu_j^{\text{embedded}}}\right \rvert.
\end{align}
Inserting unity $\sum_{i}^{\text{isolated}}\ket{\nu_i} \bra{\nu_i}$, we obtain the equivalent formulation
\begin{align}
    \left\lvert \sum \limits_{ij} \braket{F|\nu_i} \braket{\nu_i|\nu_j^{\text{embedded}}}\right\rvert.
\end{align}
A larger value implies larger scattering into vibrational modes, i.e., a weaker zero-phonon line.
Overall, a decent agreement can be observed with $\mathcal{S}^{\text{VC}}$ (compare to Fig.~\ref{fig:analysis}).
However, especially the established emitters Terrylene and DBT perform considerably worse than in the vibronic coupling entropy.
Given that those are established candidates, it seems reasonable to conclude that the entropic estimate of the scattering in $\mathcal{S}^{\text{VC}}$ is partially correcting for the missing Debye-Waller term and thus provides an overall better description of the involved physics.
Recall that relaxation of the entire supercell with TDDFT is computationally prohibitive. Future work will explore the agreement with embedding approaches such as ONIOM \cite{zirkelbach2022high}.

\section{Huang-Rhys factors -- ORCA vs. ASE/MACE-OFF}\label{app:hr-comparison}
\begin{figure*}
    \centering
    \subfigure[\label{fig:perylene_hr_spectra}]{\includegraphics[width=0.47\linewidth]{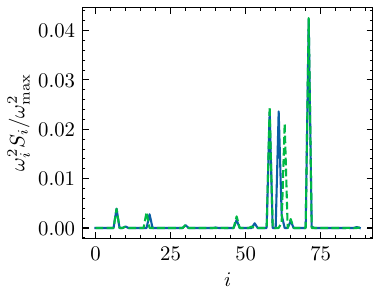}}
    \subfigure[  \label{fig:chiral_hr_spectra}]{\includegraphics[width=0.49\linewidth]{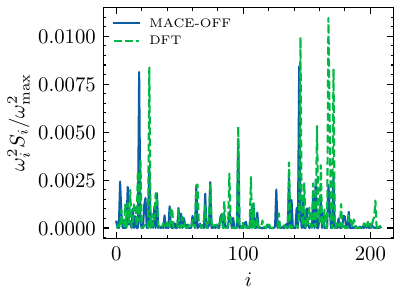}}
    \caption{\textbf{Benchmark DFT vs MACE-OFF:} Weighted Huang-Rhys factors are calculated using the same GS and excited-state configurations (obtained with DFT) but different vibrational modes.
    \textbf{(a) Perylene:} The planar structure and simplicity of perylene is accurately captured by MACE-OFF.
    Merely observable are slight switches in the mode-ordering and minor deviations in the strength of Huang-Rhys factors.
    \textbf{(b) Chiral emitter:} While the overall structure is well preserved, we can identify deviations in the low- and high-frequency domain.
    Notable normal-modes in the low-frequency domain involve spring-like oscillations which are mediated via the weak van-der-Waals interaction along the helical axis.
    Notable high-energy modes are dominated by hydrogen oscillations.
    It is not surprising that MACE-OFF performs worse in predicting weaker bonds, yet the overall structure is well preserve which indicates that our conclusions remain valuable even for more complex emitter-host pairs.}
    \label{fig:dft_vs_mace}
\end{figure*}
A vibrational system settled in equilibrium will be pushed out of the latter after an electronic transition of the system.
This is a natural consequence of the (nearly) instantaneous changes in electronic structure, i.e., the potential-energy surface, and the following forces acting on the nuclei.
This effect is known as vibronic coupling and often described in the Franck-Condon approximation which assumes indeed an instantaneous transition and harmonic potentials.
The displacement between the two vibrational harmonic potentials is often quantified in the dimensionless Huang-Rhys factors.

In a real system, many more than just a single vibration exist and the Huang-Rhys factors are defined per eigenmode.
The latter are routinely calculated by electronic structure codes such as ORCA and ASE.
Vibrational eigenmodes $u_k$ and the corresponding frequencies $\omega_l$ are obtained from the diagonalization of the Hessian
\begin{equation}
\left[\textbf{M}^{-1/2}\textbf{K}\textbf{M}^{-1/2}\right]\mathbf{u}_k = \omega_k^2 \mathbf{u}_k.
\end{equation}
If we assume the same harmonic potential in the ground and excited state, which is often a decent approximation for the relatively rigid molecular systems investigated here, then the shift in equilibrium position $\bm{\Delta} = \textbf{M}^{1/2}(\textbf{R}_{S0}-\textbf{R}_{S2})$ between the harmonics defines the strength of interaction.
Finally, the Huang-Rhys factor $S_k = \omega_k \Delta_k^2/2$ is calculated from
\begin{align}\label{eq:hrfactor}
    S_k \equiv& \frac{\omega_k}{2} ((\textbf{R}_{S0}-\textbf{R}_{S1}) \cdot \sqrt{\textbf{M}} \cdot \textbf{u}_k)^2.
\end{align}
We then compare the frequency-weighted Huang-Rhys factors $\omega_k^2 S_k/ \omega_{\text{max}}^2$ against our alternative figures of merit.

In practice, special attention has to be given to the normalization of the eigenmodes, especially their mass weighting.
While the eigenmodes provided by ASE are normalized modes weighted by the inverse of the mass $(\sqrt{\textbf{M}})^{-1}\cdot \frac{\textbf{u}_{k}}{\left \lvert \textbf{u}_{k} \right \rvert }$, ORCA provides mass-weighted modes that are normalized 
$ \frac{(\sqrt{\textbf{M}})^{-1}\cdot\textbf{u}_{k}}{\left \lvert (\sqrt{\textbf{M}})^{-1}\cdot \textbf{u}_{k} \right \rvert }$.
In order to utilize the normal modes in the basis-set expansion and the calculation of the Huang-Rhys factors, we require normalized modes in the form $\sqrt{\textbf{M}}\cdot \frac{\textbf{u}_{k}}{\left \lvert \textbf{u}_{k} \right \rvert }$.
Therefore we re-weight the modes obtained with ASE with $\textbf{M}$ and the modes obtained from ORCA with $\frac{\textbf{M}}{\lvert \sqrt{\textbf{M}}\rvert}$.
Figures~\ref{fig:perylene_hr_spectra} and~\ref{fig:chiral_hr_spectra} demonstrate the overall agreement between full TDDFT using ORCA and the usage of MACE-OFF with ASE.
The normal modes $\vert \nu_i\rangle$ are obtained from ASE by taking the exported modes and weighting them with $\sqrt{\textbf{M}}$, i.e., the  $\vert \nu_i\rangle$ provide an orthonormal basis.


We recently became aware of Ref.~\cite{sharma2025accelerating} which explores the usage of foundation machine-learning models for the prediction of vibronic coupling in point-defects systems. 
While not overlapping with the objectives of this work, it further demonstrates the utility of foundation models for optical characterization.

\section{SOAP-based Similarity}\label{app:soapsimi} 

SOAP descriptors were generated with the following parameters: A cut-off length of $r_{\text{cut}}=4$, maximum radial basis functions $n_{\text{max}}=8$, and maximum spherical harmonics degree $l_{\text{max}}=6$. Moreover, the \textit{inner average} of the descriptor is calculated, i.e., the averaging over atomic sites was performed before summing over the magnetic quantum number. This produces a new descriptor with a size irrespective of the number of atoms in the original structure. As a similarity measure between averaged descriptors, a normalized linear kernel was used
\begin{equation}
    K(a,b)=\frac{a\cdot b}{|a||b|}\,,
\end{equation}
where $a$ and $b$ denote two different descriptors.\\

\begin{figure*}[!t]
    \centering
    \includegraphics[width=\linewidth]{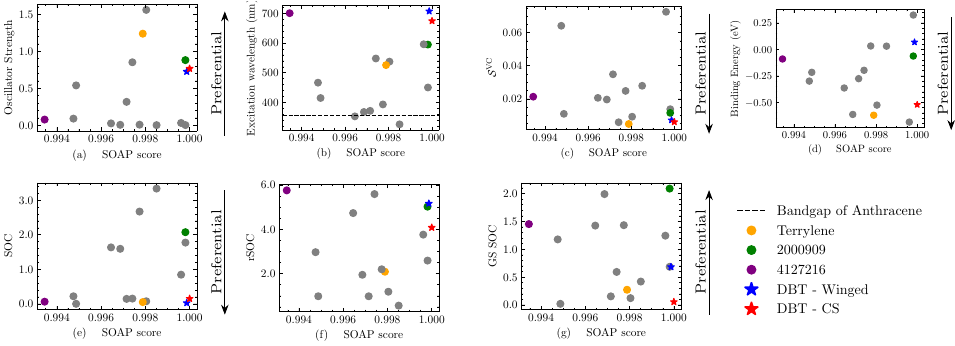}
    \caption{\textbf{Microscopic Analysis vs SOAP averages:} Illustration of the data identical to Fig.~\ref{fig:analysis} with SOAP-based similarities defined in the text.
    The lack of global structural information reduces the degree of correlation between the similarity descriptor and the various relevant quantities, especially wavelength and oscillator strength.
    }
    \label{fig:datapanel_soap}
\end{figure*}

Fig.~\ref{fig:datapanel_soap} illustrates the same results as in Fig.~\ref{fig:analysis} but plotted over the SOAP-based  similarities (see Tab.~\ref{tab:table_emitters}) in contrast to the previously used Tanimoto index.
Two differences stand out:
(i) Average SOAPs lack global information, e.g. the clear correlation between molecular size and emission wavelength (particle in a box physics) that we observe with the Tanimoto index is lost. The focus is stronger on average atomic distances, which results in a broader distribution and visually weaker correlation with physical observables.
(ii) Connected to (i), Hexa-peri-hexabenzo[7]helicene (4127216) is placed at the leftmost extreme of the evaluated set of emitters, despite comparable values with DBT in many relevant observables.
The global information encoded in Tanimoto indices via Morgan fingerprints is therefore more advantageous for our application than average SOAPs.
A combined treatment, balancing local atomic (e.g. SOAP or MACE) and global chemical structure, will be the target of future studies.

\begin{table}
\caption{\label{tab:table_emitters}\textbf{SOAP Similarity:} Similarity scores of average SOAP descriptors of investigated emitters,  with DBT - CS as reference structure. A normalized linear kernel is used as similarity measure.}
\resizebox{0.5\textwidth}{!}{
\begin{tabular}{lcc}
 \toprule \toprule
 Name/COD ID&SOAP similarity score
\\ \midrule
DBT - CS/-& $1.0$\\
DBT - Winged/-& $0.9999$\\
 Hexa-peri-hexabenzo[7]helicene/4127216&$0.9934$ \\  Tetrabenzo[\textit{de},\textit{hi},\textit{op},\textit{st}]pentacene/2000909&$0.9998$ \\
 Terrylene/-&$0.9979$ \\
 Perylene/-&$0.9949$ \\
BDPB/-&$0.9974$  \\
Diindenozethrene/1555531&$0.9996$  \\
-/4062916&$0.9985$ \\
Benzo[\textit{ghi}]perylene/1554212&$0.9972$  \\
naphtho[1,2-i]pentahelicene/7155241&$0.9977$ \\
DPNP/-&$0.9980$ \\
Diacenaphtheno[7,8-b:$7^{1}$,$8^{1}$-d]thiophene/2203348&$0.9948$ \\
-/4118733&$0.9969$ \\
Dibenzo[g,p]chrysene/4107152&$0.9964$ \\
Pentabenzo[a,d,g,j,m]coronene/4037514&$0.9998$ \\
\bottomrule
\bottomrule
\end{tabular}
}
\end{table}


\section{Embedding Strategy}\label{app:embedding}
The embedding strategy is detailed in algorithm~\ref{alg:cap}.
We begin by placing the emitter into the center of mass of the host.
A large number of individual trial-insertions is then performed.
Each time, the emitter is randomly translated and rotated, ensuring in the end that the molecule remains within the supercell.
Next, we mark all atoms that are in close proximity (here a cut-off of 1~\AA~ is selected) with the embedded molecule.
Finally, all molecules for which some atoms are marked as ``close" are then removed from the combined system.

Subsequent to this embedding strategy follows the relaxation of the combined system using the ASE-internal LBFGS algorithm using the MACE-OFF potential and a maximally allowed force of 0.01 $\text{eV}/$\AA.
The same relaxation is performed for the supercell and emitter separately.
Finally, the formation (binding) energy is calculated, and only the energetically most favorable configuration is retained for all following studies.

\begin{algorithm*}
\caption{Embedding algorithm}\label{alg:cap}
\begin{algorithmic}
\State host\_molecules $\gets \Call{IdentifyMolecules}{\text{Host}}$ \Comment{Return a list of molecular indices for each atom}
\State $(v_a, v_b, v_c) \gets \text{Host}\texttt{.get\_cell\_lengths}()$
\State $\text{Emitter} \gets \text{Emitter}.\texttt{translate}(\text{Host}.\texttt{get\_center\_of\_mass}())$ \Comment{Place emitter at host center of mass}
\\
\For{trial in \texttt{range}($N_{\text{max}}$)}
\\
\State $\vec{r}_{\text{axis}} \gets \text{random unit vector}$
\State $\theta \gets \text{random angle in } [0^\circ, 359^\circ)$
\State Emitter $\gets \text{Emitter}\texttt{.rotate}(\theta,\vec{r}_{\text{axis}})$\Comment{Rotate emitter}
\\
\State $\vec{r}_{\text{shift}} \gets 0.05 \cdot [v_a \cdot \texttt{rand}(-1,1), v_b \cdot \texttt{rand}(-1,1), v_c \cdot \texttt{rand}(-1,1)]$
\State Emitter $\gets \text{Emitter}.\texttt{translate}(\vec{r}_{\text{shift}})$ \Comment{Translate emitter}
\\
\State \text{overlapping\_atoms} $\gets \texttt{[]}$
\For{each atom $a_1$ in Host}
    \For{each atom $a_2$ in Emitter}
        \If{$\|a_2\texttt{.position} - a_1\texttt{.position}\| < 1$}
            \State $\text{overlapping\_atoms}\texttt{.append}(a_1\texttt{.index})$ \Comment{Mark host atom as overlapping}
        \EndIf
    \EndFor
\EndFor
\State overlapping\_molecules $\gets \{\text{host\_molecules}[i] \mid i \in \text{overlapping\_atoms} \}$ \Comment{Get host molecules containing overlapping atoms}

\State \text{indices\_to\_delete} $\gets \{ i \mid \text{host\_molecules}[i] \in \text{overlapping\_molecules} \}$
\State \texttt{delete}(\text{Host}[\text{indices\_to\_delete}])\Comment{Delete overlapping host molecules}
\\
\State $\text{host\_max} \gets [\max(x), \max(y), \max(z)]$ from \text{Host}\texttt{.get\_positions}()
\State $\text{host\_min} \gets [\min(x), \min(y), \min(z)]$ from \text{Host}\texttt{.get\_positions}()
\\
\State $\text{emitter\_max} \gets [\max(x), \max(y), \max(z)]$ from \text{Emitter}\texttt{.get\_positions}()
\State $\text{emitter\_min} \gets [\min(x), \min(y), \min(z)]$ from \text{Emitter}\texttt{.get\_positions}()
\\
\If{\texttt{all}(\text{emitter\_min} $>$ \text{host\_min}) and \texttt{all}(\text{emitter\_max} $<$ \text{host\_max)}} \Comment{Ensure emitter is contained within host
}
\State \texttt{Save}(Host + Emitter)
\EndIf
\EndFor
\end{algorithmic}
\end{algorithm*}

\begin{algorithm*}
    \begin{algorithmic}
        \Function{IdentifyMolecules}{Atoms}
    \State cutoffs $\gets \texttt{ase.neighborlist.natural\_cutoffs}(\text{Atoms})$
    \State $\text{NL} \gets \texttt{ase.neighborlist.NeighborList}(\text{cutoffs},\texttt{self\_interaction=False, bothways=True})$
    \State $\text{NL}\texttt{.update}(\text{Atoms})$
    \State M $\gets \text{NL}\texttt{.get\_connectivity\_matrix}()$
    \State  $\_, \text{molecule\_list} \gets \texttt{scipy.sparse.csgraph.connected\_components}(\text{M})$ 
    \State \Return \text{molecule\_list}
\EndFunction
\end{algorithmic}
\end{algorithm*}

\section{DBT geometry}
\label{app:dbt} 
\begin{figure}[h]
  \centering
  \includegraphics[width=\columnwidth]{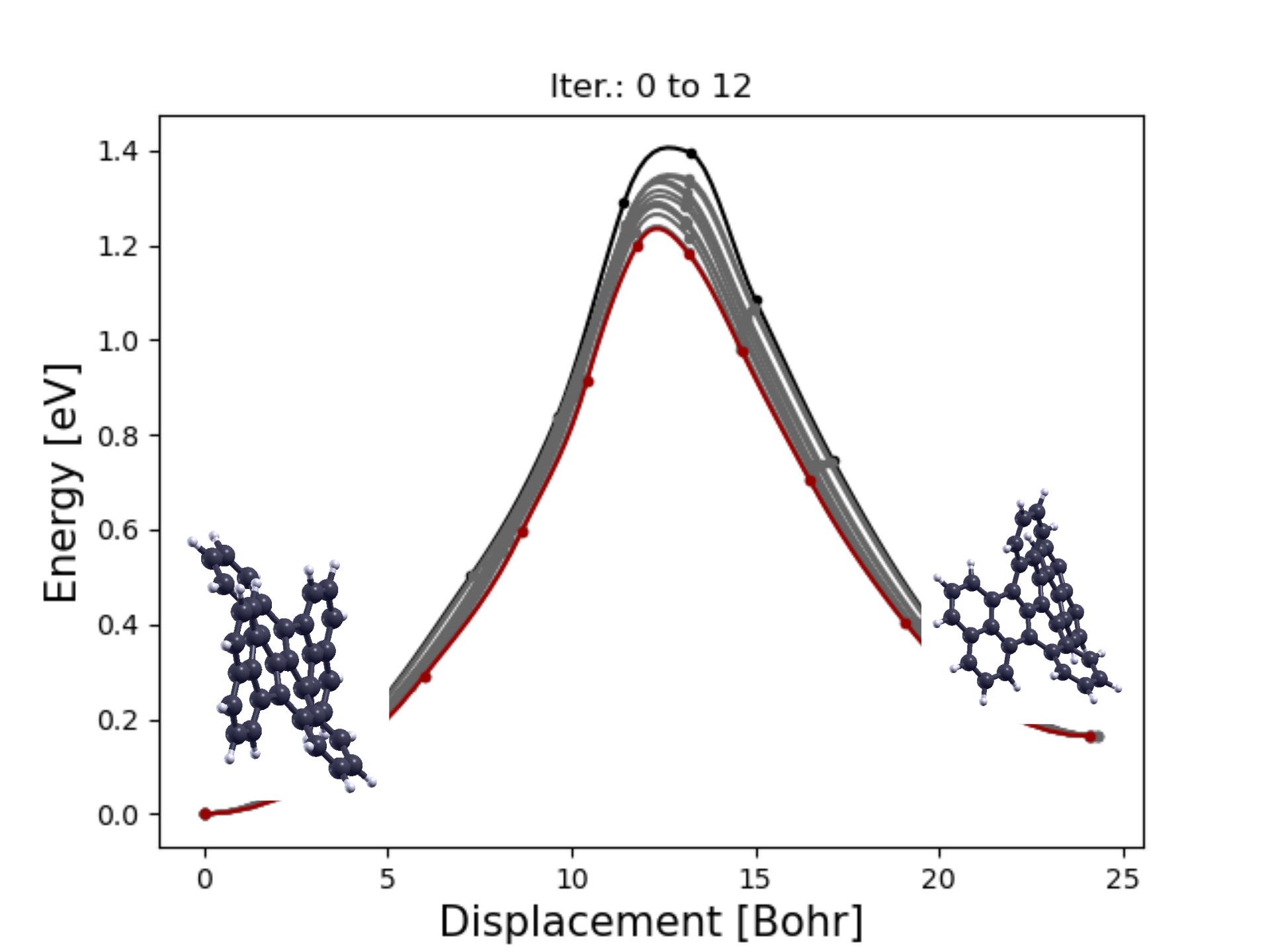}
  \caption{\textbf{Nudged-elastic band calculations}: 
  Relaxation of the nudged-elastic band for the lowest energy path between both conformations (max force $< 0.0025$). The centrosymmetric (left of barrier) and winged (right of barrier) structure are separated by a barrier with an approximate height of 1.2~eV. It is therefore unlikely that thermal fluctuations would result in a spontaneous conversion.
  }
  \label{fig:neb}
\end{figure}

A geometry optimization from the 2D structure converges into two possible minima: (i) the centrosymmetric structure, or (ii) a winged conformation (see Fig.~\ref{fig:neb}).
Both configurations have the respective energies
$E_{\text{B3LYP+D3}}^{\text{def2-TZVPD}}(R_{\text{CS}}) = -1459.918888$~H and
$E_{\text{B3LYP+D3}}^{\text{def2-TZVPD}}(R_{\text{winged}}) = -1459.912846$~H, with an energy difference between the structures of $\Delta E_{\text{B3LYP+D3}}^{\text{def2-TZVPD}} = 0.1644$~eV in favor of the centrosymmetric structure.
Respective vertical excitation energies for absorption are
$\lambda_{\text{B3LYP}}^{\text{def2-TZVPD}}(R_{\text{CS}})=674.8$~nm
and 
$\lambda_{\text{B3LYP}}^{\text{def2-TZVPD}}(R_{\text{winged}})=706.7$~nm, with emission energies 
$\lambda_{\text{B3LYP}}^{\text{def2-TZVPD}}(R_{\text{CS}})=706.3$~nm
$\lambda_{\text{B3LYP}}^{\text{def2-TZVPD}}(R_{\text{winged}})=739.9$~nm.

Using a smaller basis and ignoring dispersive interactions results in direct relaxation into the centrosymmetric structure, with respective energies
$E_{\text{B3LYP}}^{\text{6-31G}}(R_{\text{CS}}) = -1458.911435$~H
$E_{\text{B3LYP}}^{\text{6-31G}}(R_{\text{winged}}) = -1458.905947$~H
and energy difference $\Delta E_{\text{B3LYP}}^{\text{6-31G}} = 0.1493$~eV.
The most reliable predictions can be obtained when combining different perspectives, i.e., using an ensemble of different functionals.
Therefore, we decided to perform the same evaluation using the meta-GGA functional r2SCAN in combination with the dispersion correction D4. 
This combination is able to self-consistently treat short- and medium-ranged dispersion and perturbatively accounts for long-range dispersion, at the cost of self-interaction errors that are partially removed with hybrids such as B3LYP.
Despite the conceptual differences, our results remain qualitatively consistent where
$E_{\text{r2SCAN+D4}}^{\text{def2-TZVPD}}(R_{\text{CS}}) = -1459.896092$~H
$E_{\text{r2SCAN+D4}}^{\text{def2-TZVPD}}(R_{\text{winged}}) = -1459.860686$~H
and $\Delta E_{\text{r2SCAN+D4}}^{\text{def2-TZVPD}} = 0.9634$~eV.

We observe a consistent trend: every geometry optimization that accounts for dispersive interactions will favor the winged structure during relaxation, despite its elevated energy compared to the centrosymmetric structure.
We performed nudged-elastic band calculations to identify the effective energetic barrier between the two conformations.
As illustrated in Fig.~\ref{fig:neb}, we identify a barrier of approximately 1.2 eV, which is unlikely to be overcome due to thermal fluctuations.

Despite a clear energetic preference for the centrosymmetric structure, recent experimental results (DBT in anthracene) measured a substantial linear component to the Stark shift.\cite{schadler2019electrical}
The following DFT calculations demonstrate that a winged structure features consistent Stark coefficients with the experiment, which might hint at the co-presence of two stable conformers.



Following first and second order perturbation theory, the Stark shift is given by 
\begin{align*}
    \Delta E = -\boldsymbol\mu \cdot \textbf{E} - \frac{1}{2} \textbf{E} \cdot  \boldsymbol\alpha \cdot \textbf{E} + O(E^3).
\end{align*}
DBT is randomly oriented with respect to the external field, which we model by using the magnitude for the dipole moment and the average values for the polarizability.
Excitation energies are then approximately shifted by 
\begin{align*}
    h\Delta \nu &\approx \vert \delta\mu \vert E - \frac{1}{2} \delta\alpha E^2\\
\end{align*}
where  $\delta$ indicates the difference between ground and relevant excited state.
Calculating permanent dipoles and polarizability is a more complex endeavor for the excited states (with ORCA).
We have decided to base our theoretical discussion on the ground-state quantities, as arguments about centrosymmetry apply equivalently for the excited state.
Extracted linear and quadratic parameters for the shift
\begin{align*}
    \Delta \nu &\approx a \vert E\vert + b \vert E\vert^2
\end{align*}
are then compared to experimentally fitted values (see Tab.~\ref{tab:starkshift}).

\begin{table*}
\caption{
\textbf{Dependence of the Stark shift on geometry, functional, and basis set:}
A distinct qualitative difference in permanent dipole and linear Stark shift can be found between the centrosymmetric and winged structure.
All geometries are optimized in the local minimum with the respective functionals and basis sets.
Please note that our calculations to do not account for changes in dipole and polarizability of the excited state.
Despite this limitation, the qualitatively agreement between winged structure and experiment indicate that the winged structure might be experimentally relevant.
}
    \centering
    \resizebox{\textwidth}{!}{
    \begin{tabular}{lccccccccc}
     \toprule \toprule
        Functional & Basis & Structure & $\vert \mu_0\vert$ (a.u.) & $\alpha_0$ (a.u.) & a ($\text{MHz}~\text{kV}^{-1}~\text{cm}$) & b ($\text{MHz}~\text{kV}^{-2}~\text{cm}^2$) \\
        \toprule 
        B3LYP & 6-31G & CS & 0.0002  & 536.13 &  0.242 & -0.0667 \\
        B3LYP D3 & def2-TZVPD & CS & 0.0000 & 598.76 & 0.038 & -0.0745 \\
        r2SCAN D4 & def2-TZVPD & CS & 0.0001 & 608.63 & 0.111 & -0.0757 \\
        \midrule
        PBE D3 & def2-TZVPD & winged & 0.0458 & 644.28 &  58.58 & -0.0801\\
        B3LYP D3 & def2-TZVPD & winged & 0.0616 & 612.15 & 78.84 & -0.0761\\
        r2SCAN D4 & def2-TZVPD & winged & 0.0668 & 617.43 & 85.44 & -0.0768 \\
        \midrule
        Experiment\cite{schadler2019electrical} &  & - & & & 300 & -0.15\\
        \bottomrule
        \bottomrule
    \end{tabular}
    }
    \label{tab:starkshift}
\end{table*}




\section{Investigated Molecules}
\begin{figure} 
    \centering
    \subfigure[]{\includegraphics[width=0.15\textwidth]{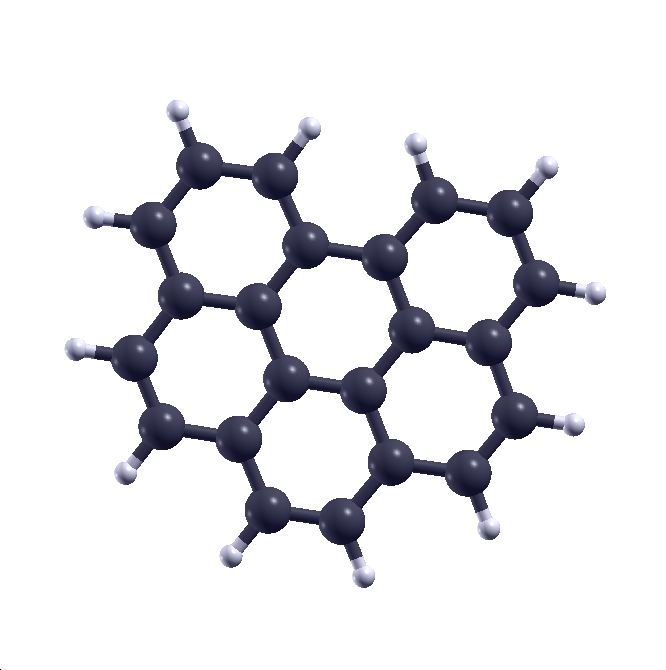}}
    \subfigure[]{\includegraphics[width=0.15\textwidth]{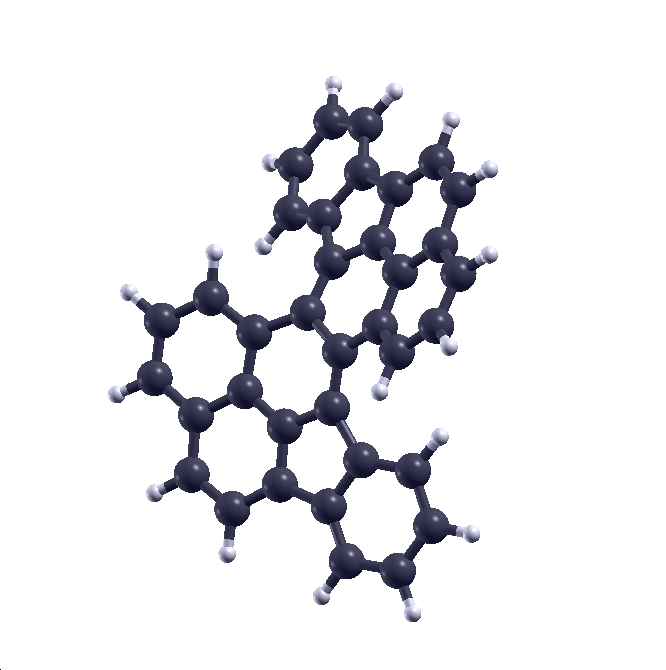}} 
    \subfigure[]{\includegraphics[width=0.15\textwidth]{2000909_go.png}}
    \subfigure[]{\includegraphics[width=0.15\textwidth]{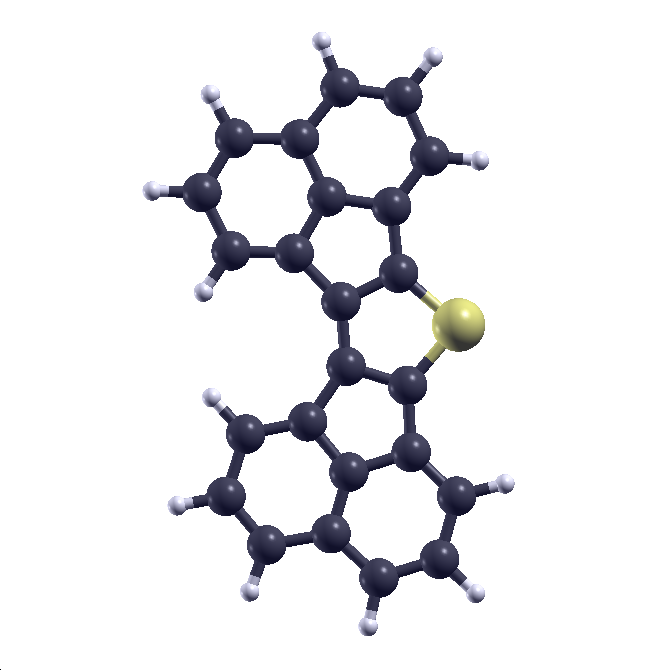}}
    \subfigure[]{\includegraphics[width=0.15\textwidth]{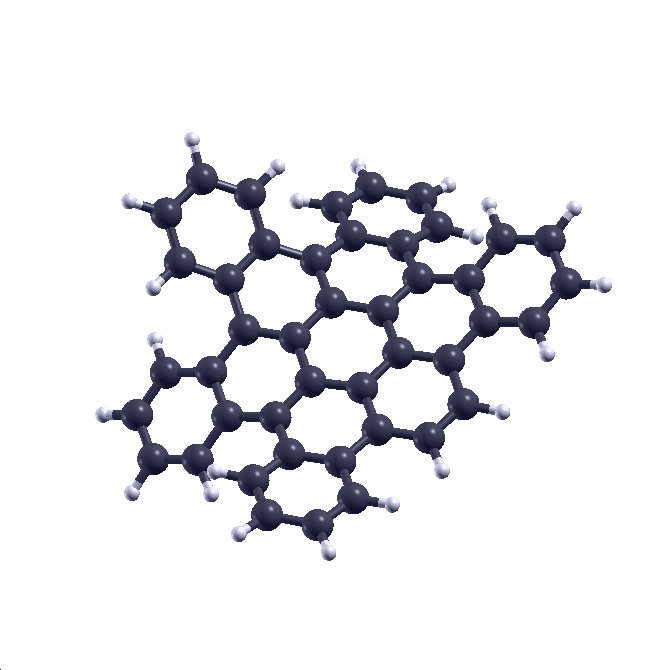}}
    \subfigure[]{\includegraphics[width=0.15\textwidth]{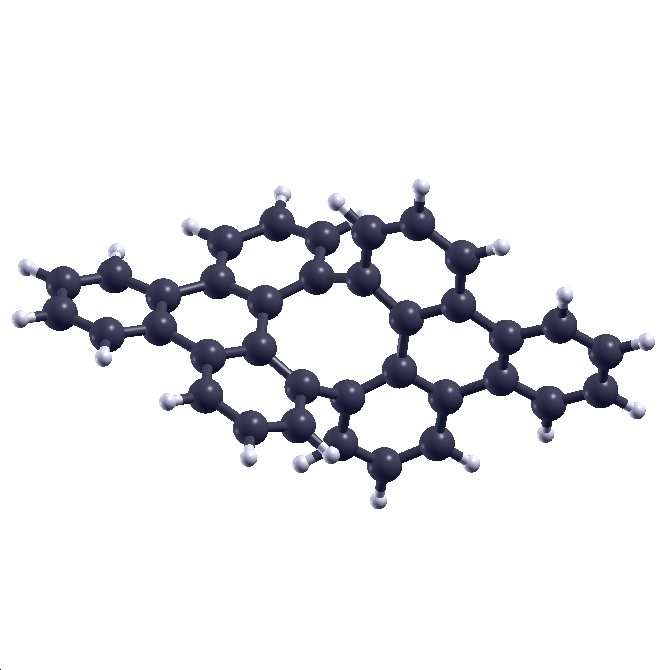}}
    \subfigure[]{\includegraphics[width=0.15\textwidth]{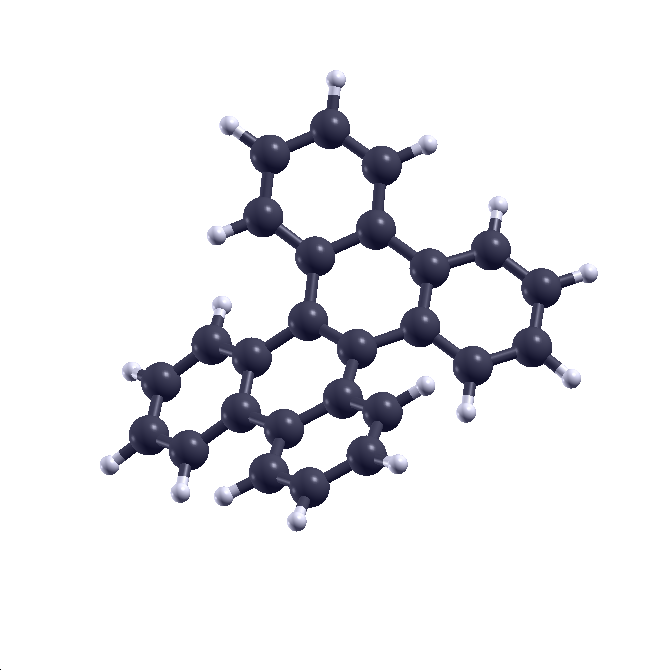}}
    \subfigure[]{\includegraphics[width=0.15\textwidth]{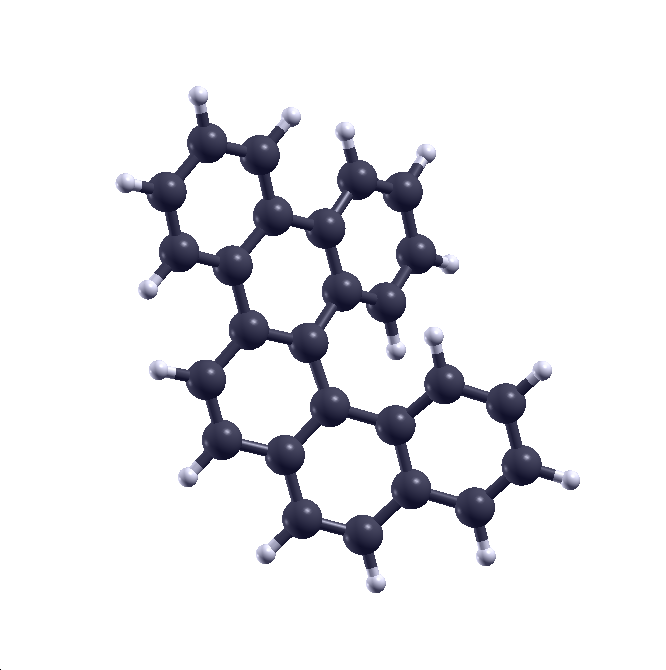}}
    \subfigure[]{\includegraphics[width=0.15\textwidth]{4127216_go.png}}
    \subfigure[]{\includegraphics[width=0.15\textwidth]{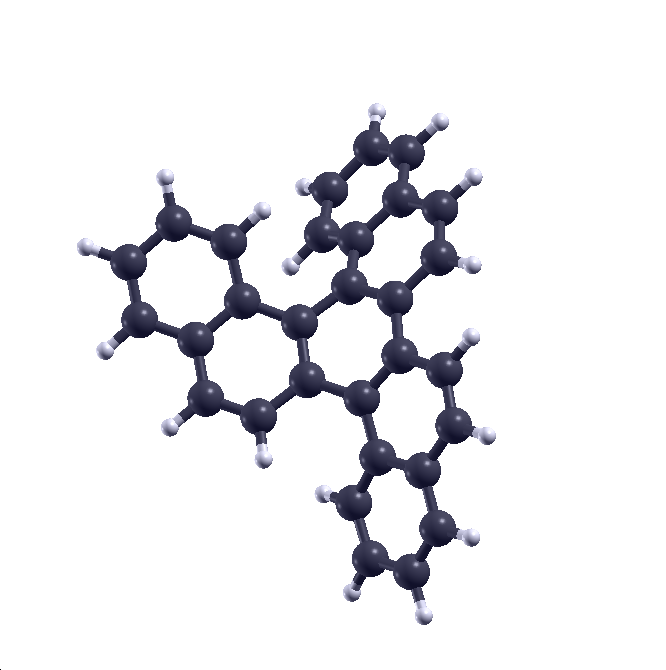}}
    \subfigure[]{\includegraphics[width=0.15\textwidth]{BDPB_go.png}}
    \subfigure[]{\includegraphics[width=0.15\textwidth]{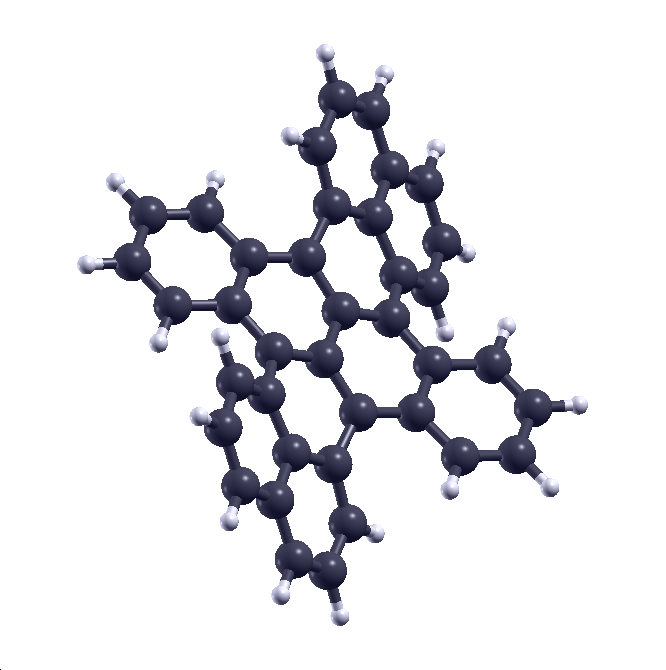}}
    \subfigure[]{\includegraphics[width=0.15\textwidth]{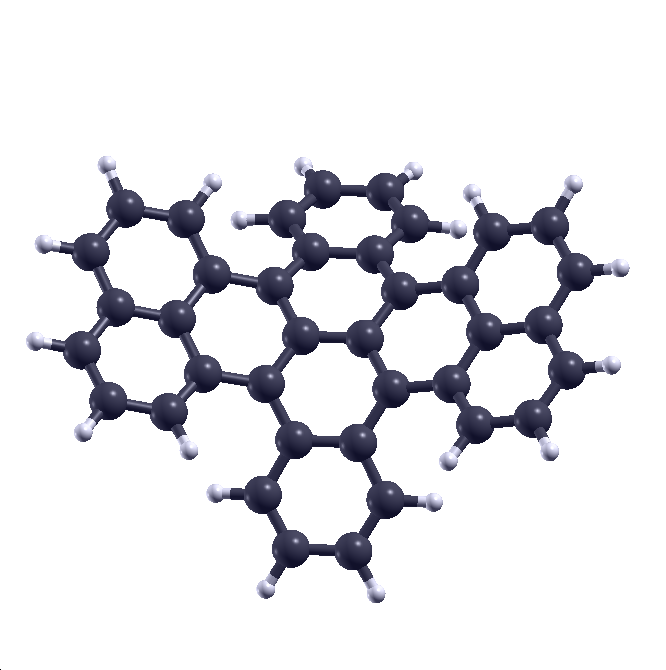}}
    \subfigure[]{\includegraphics[width=0.15\textwidth]{DPNP_go.png}}
    \subfigure[]{\includegraphics[width=0.15\textwidth]{Perylene_go.png}}
    \subfigure[]{\includegraphics[width=0.15\textwidth]{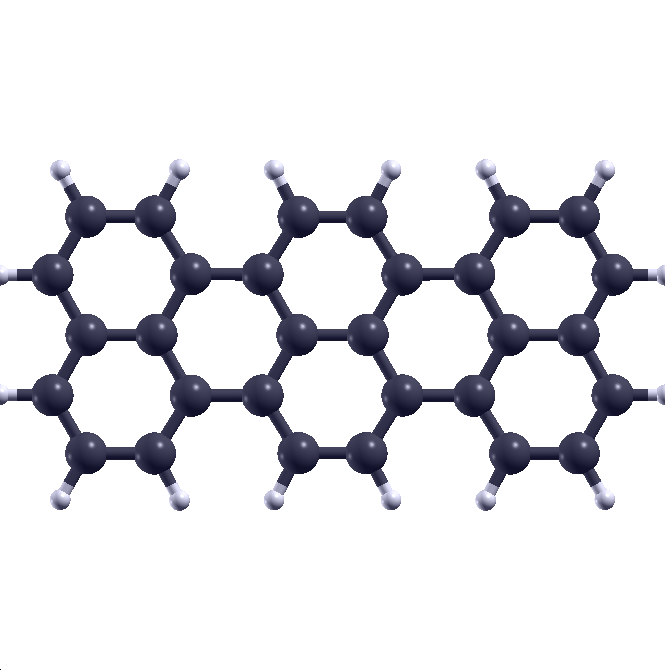}}
    \caption{\textbf{Emitter candidates:}
    \textbf{(a)} CID: 1554212,
    \textbf{(b)} CID: 1555531,
    \textbf{(c)} CID: 2000909,
    \textbf{(d)} CID: 2203348,
    \textbf{(e)} CID: 4037514,
    \textbf{(f)} CID: 4062916,
    \textbf{(g)} CID: 4107152,
    \textbf{(h)} CID: 4118733,
    \textbf{(i)} CID: 4127216,
    \textbf{(j)} CID: 7155241,
    \textbf{(k)} BDPB,
    \textbf{(l)} DBT (CS), 
    \textbf{(m)} DBT (winged),
    \textbf{(n)} DPNP,
    \textbf{(o)} Perrylene and
    \textbf{(p)} Terrylene.
    }
    \label{fig:allstructures}
\end{figure}

Fig.~\ref{fig:allstructures} illustrates the structure of all organic molecules investigated in this work as potential SPE candidates in anthracene.

\begin{figure}
  \centering
  {\includegraphics[width=0.49\textwidth]{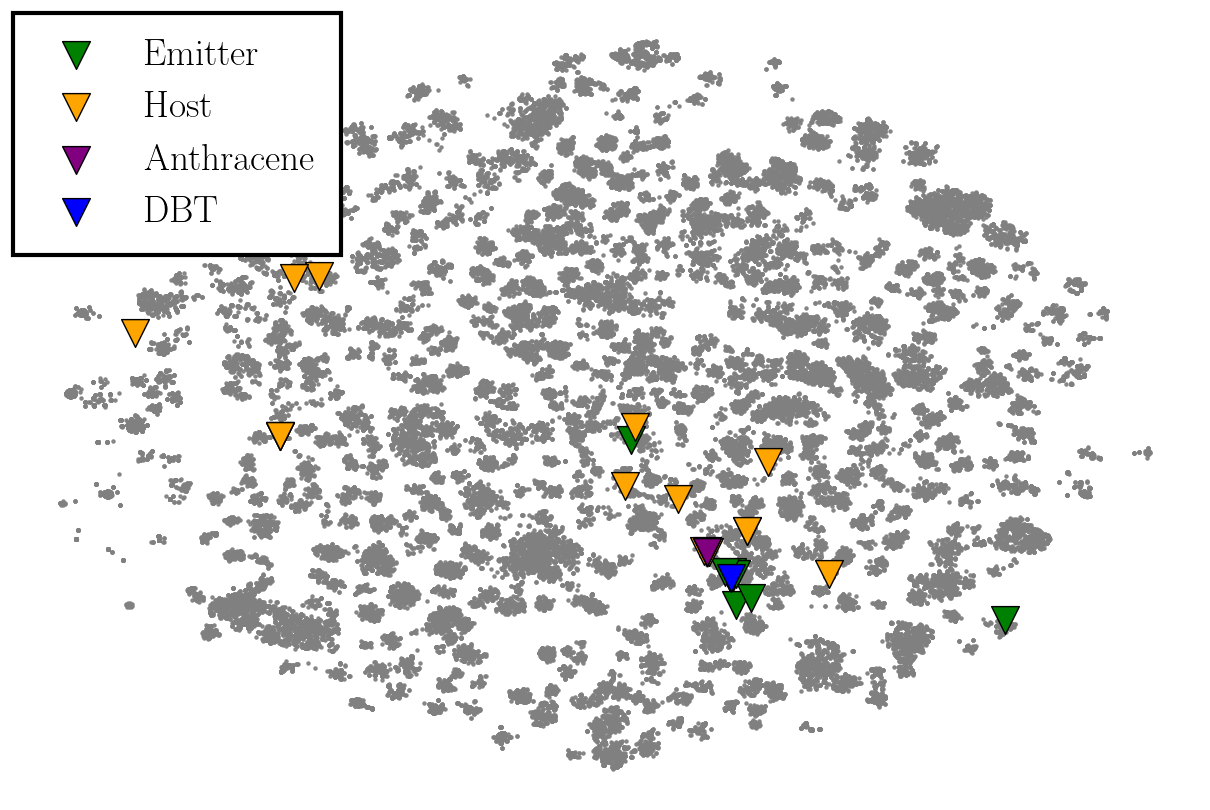}}
\caption{
\textbf{Global structure of dataset - Emitters \& Hosts:} t-SNE plot showing the distribution of molecular data. Known emitters (hosts) are indicated by green (orange) triangles, and the reference DBT (anthracene) is highlighted by a blue (purple) triangle.} 
\end{figure}

\bibliography{main}

\end{document}